\documentstyle[twocolumn,psfig,aps]{revtex}
\draft	

\begin{document}
\twocolumn[\hsize\textwidth\columnwidth\hsize\csname @twocolumnfalse\endcsname
\title{Dynamics of Ordering of Heisenberg Spins with Torque --- Nonconserved 
Case. I}
\author{Jayajit Das\cite{JAY} and Madan Rao\cite{MAD}}

\address{Institute of Mathematical Sciences, Taramani, Chennai (Madras) 
600113, India}

\date{\today}

\maketitle

\begin{abstract}
We study the dynamics of ordering of a nonconserved Heisenberg magnet. The
dynamics consists of two parts --- an irreversible dissipation into a heat
bath and a reversible precession induced by a torque due to the local
molecular field. For quenches to zero temperature, we provide convincing
arguments, both numerically (Langevin simulation) and analytically
(approximate closure scheme due to Mazenko), that the torque is irrelevant
at late times. We subject the Mazenko closure scheme to systematic 
numerical tests. Such an analysis, carried out for the first time on a
vector order parameter, shows that the closure scheme performs respectably 
well. For quenches to $T_c$, we show, to ${\cal O}(\epsilon^2)$, that
the torque is irrelevant at the Wilson-Fisher fixed point. 
\end{abstract}

]
\vskip1.0in

\section{Introduction}

Interacting systems like magnets and binary fluids exhibit an
ordered configuration at low temperatures, consisting of coexisting,
symmetry broken phases. When cooled rapidly from the disordered phase at
high temperatures, such systems take a long time to establish order,
primarily because of the slow annealing of the interfaces (defects)
separating the competing domains. At late times, the system organizes
itself into a self similar spatial distribution of domains characterised
by a single diverging length scale which typically grows algebraically in
time $L(t) \sim t^{1/z}$. This spatial distribution of domains is
reflected in the scale dependent behaviour of the correlation functions. 

In the last few years, a fairly detailed picture of the late time
behaviour of the correlation functions has emerged \cite{BRAY}. The
equal-time order parameter correlation function $C(r, t) \equiv \langle
{\vec \phi}(r, t) \cdot {\vec \phi}(0,t) \rangle$ is a measure of the
spatial distribution of the domains, and at late times is found to behave
as $f(r/L(t))$, where $L(t)$ is the distance between defects.  The
autocorrelation function, $C(0, t_{1}=0, t_{2}) \equiv \langle \,{\vec
\phi}(0, 0) \cdot {\vec \phi}(0, t_2) \rangle$, is a measure of the memory
of the initial configurations, and decays at late times as
$L(t_{2})^{-\lambda}$. The independent scaling exponents $z$ and $\lambda$
and the scaling function $f(x)$ characterise the dynamical universality
classes at the zero temperature fixed point \cite{BRAY}. 

The scaling results referred to above, have been obtained in systems
coupled to a constant temperature heat bath, where the order parameter
${\vec \phi} (r,t)$ undergoes a purely dissipative (irreversible)
dynamics.  There is no dynamics of the order parameter in the absence of
this coupling. In general however, apart from dissipating into a heat
bath, the system may have a hamiltonian dynamics of its own. Such dynamics
can be represented by a generalized Langevin equation \cite{CHAI},
\begin{eqnarray}
\frac{\partial \phi_{\alpha}(r,t)}
{\partial t} & =  &  \int_{r',t'}
\left \{ \phi_{\alpha}(r,t), \phi_{\beta}(r',t')
\right \} \, \frac {\delta F}{\delta \phi_{\beta}
(r',t')} \nonumber \\
             &    & - \, \Gamma \,( -i\nabla)^{\mu} \frac {\delta
F}{\delta \phi_{\alpha}(r,t)}\,+ \,\eta_{\alpha}(r,t)\,\, ,
\label{eq:genlang} 
\end{eqnarray}
where $\phi_{\alpha}$ is an $N$-component order parameter, $F$ is a
coarse-grained free-energy functional, $\{ \cdot , \cdot \}$ is the
Poisson bracket, and $\eta_{\alpha}$ is the noise. The first term on the
right hand side is reversible, while the second is the usual dissipative
force, with $\mu = 2$ or $0$ according as whether the order parameter is
conserved or not. The noise correlator $\langle \eta_{\alpha}(r,t) \,
\eta_{\beta}(r',t') \rangle = 2 \,\Gamma \,k_B T \,\delta_{\alpha
\beta}\,(-i\nabla)^{\mu} \,\delta (r - r') \,\delta (t - t')$ is
proportional to the temperature $T$. 

As an example let us consider the dynamics of a binary fluid given by the
generalized Langevin description Eq.\ (\ref{eq:genlang}). The relative
concentration $\phi$ of a binary fluid is advected by the velocity field
${\bf v} = {\bf g}/\rho$. The generalized Langevin equation for $\phi$
involves the reversible $\{\phi,g_{\alpha}\} \,\delta F/\delta g_{\alpha}$
which reduces to the familiar streaming term ${\bf v} \cdot {\bf \nabla}
\phi$ \cite{BRAY}. 

One might classify the dynamics according to the algebraic structure of
the Poisson brackets. For instance, the Poisson algebra of the components
of the order parameter could be of the form $\{\phi_{\alpha}, \phi_{\beta}
\} = c$, where $c$ is in general a complex number.  A recently studied
example is the dynamics of superfluid ordering of a bose gas \cite{BOSE}.
The dynamics is written in terms of a complex boson annihilation field
$\psi$ which obeys the Poisson bracket relation $\{ \psi,\psi^{*} \} = i$.
On the other hand, the Poisson algebra of the components of the order
parameter could be a Lie algebra $\{\phi_{\alpha}, \phi_{\beta} \} =
c_{\alpha \beta \gamma} \phi_{\gamma}$, where $c_{\alpha \beta \gamma}$
are the structure constants.  A common example is the dynamics of the
local magnetic moments of a Heisenberg magnet. The
components of the magnetic moments $\phi_{\alpha}$ ($\alpha = 1, 2, 3$)
satisfy the Poisson algebra,
\begin{equation}
\{ \phi_{\alpha}( r,t),\phi_{\beta}( r',t') \} = \Omega_{L} 
\,\epsilon_{\alpha \beta \gamma }\, \phi_{\gamma}(r ,t)\,\delta(r - r')\, 
\delta(t - t')\,,
\label{eq:pbrac} 
\end{equation}
where $\epsilon_{\alpha \beta \gamma}$ is the completely antisymmetric
tensor in three dimensions and $\Omega_L$ is the Larmour frequency. 

In this paper, which appears in two parts, we discuss the phase ordering
dynamics of a Heisenberg magnet in three dimensions. Part I \,of the paper
is a detailed discussion of the nonconserved model, while Part II
discusses the conserved ordering dynamics. A section wise breakup of Part
I follows. After a brief account of the dynamics of the model (Section
II), we discuss the dynamics of phase ordering following a quench to $T=0$
(Section III). A Langevin simulation and an analytical calculation using
the approximate closure scheme of Mazenko \cite{MAZENKO,BRAY} lead to the
conclusion that the extra reversible term is irrelevant at late times. We
subject the Mazenko closure to systematic numerical tests and show that it
is a fairly good approximation at late times. This section is of general
interest, since this is the first detailed numerical `test' of the Mazenko
theory for a vector order parameter. In Section IV, we investigate the
dynamics following a quench to the critical point $T_c$ and show, using a
perturbative $\epsilon$-expansion, that the reversible term is irrelevant
at the Wilson-Fisher fixed point. 
 
\section{Heisenberg Magnet and Precessional Dynamics}

The spins $\phi_{\alpha}$ ($\alpha = 1,2,3$) in a Heisenberg ferromagnet
in three dimensions experience a torque from the joint action of the
external field (if present) and the local molecular field. In response the
spins precess with a Larmour frequency $\Omega_L$ about the total magnetic
field. Coupling to various faster degrees of freedom like lattice
vibrations or electrons, causes a dissipation in the energy and an
eventual relaxation towards equilibrium. 

This dynamics follows from the generalized Langevin equation Eq.\
(\ref{eq:genlang}), and the Poisson algebra Eq.\ (\ref{eq:pbrac}),
\begin{equation}
\frac{\partial \phi_{\alpha}}{\partial t} = -\Gamma (-i \nabla)^{\mu}\frac
{\delta F}{\delta \phi_{\alpha}}\, + \,\Omega_L \,\epsilon_{\alpha \beta \gamma}
\,\phi_{\beta}\,\frac{\delta F}{\delta \phi_{\gamma}} + \,\eta_{\alpha}\,\,.
\label{eq:dye}
\end{equation}
The free-energy functional $F$ is taken to be of the Landau form,
\begin{eqnarray}
F[\vec{\phi}] & = & \int \,d^{3}x \,\left[ \,\frac{\sigma}{2} \,(\nabla\vec{\phi})^2 
-\,\frac{r}{2}\,(\vec{\phi} \cdot \vec{\phi})\,+\,\frac{u}{4}\,(\vec{\phi} 
\cdot \vec{\phi})^2 \, \right]
\label{eq:landau} 
\end{eqnarray}
The second term in Eq.\ (\ref{eq:dye}) is clearly the torque ${\vec M}
\times {\vec H}$ where ${\vec H} \equiv \delta F/\delta {\vec \phi}$ is the
local molecular field. 

The free-energy functional is rotationally invariant in spin space and so
the Poisson bracket term conserves the total spin. If the dissipation
arises from spin-spin interactions, then this will conserve the total spin
too and so $\mu$ should be taken to be $2$. If however the dissipation is
a consequence of spin-lattice or spin-orbit interactions, then the total
spin will not be conserved and so $\mu =0$. In Part I we will consider the
case of nonconserved dynamics, leaving the conserved dynamics for Part II. 

Since the noise correlator is proportional to temperature, we may drop it
in our discussion of zero temperature quenches.  We then scale space ${\bf
x}$, time $t$ and the order parameter $\vec{\phi}$ as
\[ {\bf x } \rightarrow \sqrt{\frac{r}{\sigma}}\,{\bf x} , \: \:  t
\rightarrow \Gamma r t, \: \: \vec \phi \rightarrow \sqrt{\frac{r}{u}}
\vec \phi \]
to obtain the equation of motion in dimensionless form,
\begin{equation}
\frac{\partial \vec{\phi}}{\partial t} =  \nabla^2 \vec{\phi} \, + \, 
\vec{\phi} - \, \left(\vec{\phi} \cdot \vec{\phi}\right)\, \vec{\phi} \,
 + \,g \, \left(\vec{\phi} \times \nabla^2\vec{\phi}\right)
\label{eq:dyeq2}
\end{equation}
The dimensionless parameter $g = ({\Omega_{L}}/{\Gamma})\sqrt{r/u}$ is the
ratio of the precession frequency to the relaxation rate. To get a feel for
the values assumed by $g$, let us set $\Omega_L \sim 10^{7}$\,Hz, $\Gamma
\sim 10^6 - 10^{10}$\,Hz, which gives a range of $g \sim 10^{-3} - 10$. 

\section{Phase Ordering Dynamics\,: Quenches to $T = 0$}

Let us now prepare the system initially in the paramagnetic phase and quench
to zero temperature. We study the time evolution of the spin 
configurations as they evolve according to Eq.\ (\ref{eq:dyeq2}). We 
calculate the equal time correlator,
\begin{equation}
 C({\bf r} ,t ) \equiv \langle {\vec \phi({\bf r} ,t)} \cdot {\vec \phi({\bf 
r+x},t)}\rangle\,,
\label{eq:eqcorr}
\end{equation}
and the autocorrelator,
\begin{equation}
C({\bf 0}, t_1=0, t_2) \equiv \langle \vec \phi({\bf r}, t_1=0) \cdot \vec 
\phi({\bf r}, t_2) \rangle \,,
\label{eq:autocorr}
\end{equation}
where the angular brackets are averages over the random initial
conditions and space. At late times these correlators should attain their 
scaling form
\begin{equation}
C({\bf r}, t) \sim f(r/L(t))
\label{eq:scorr}
\end{equation}
\begin{equation}
C({\bf 0}, t_1=0, t_2) \sim L(t_2)^{-\lambda}\,.
\label{eq:sauto}
\end{equation}
The length scale $L(t)$ is a measure of the distance between defects and 
grows with time as $L(t) \sim t^{1/z}$. We compute the scaling function 
$f(x)$, the growth exponent $z$ and the autocorrelation exponent $\lambda$
by ($i$) simulating the Langevin Eq.\ (\ref{eq:dyeq2}) and ($ii$) 
Mazenko's closure approximation.

\subsection{Langevin Simulation}  	

We discretize Eq.\ (\ref{eq:dyeq2}) on a simple cubic lattice (with size 
$N$ ranging from $40^3$ to $60^3$) adopting an Euler scheme for the 
derivatives \cite{SR}. The space and time intervals have been chosen to 
be $\triangle x = 3$ and $\triangle t = 0.01$. With this choice of 
parameters, we have checked that the resulting coupled map does not lead 
to any instability. We have also checked that the results remain 
unchanged on slight variations of $\triangle x$ and $\triangle t$. Unless 
otherwise specified, all calculated quantities are averaged over $30$ 
uncorrelated initial configurations taken from a uniform distribution 
with zero mean. Throughout our simulation we have used periodic boundary 
conditions.

The scaling of the energy density,
\begin{equation}
\varepsilon = \frac{1}{V}\, \int d {\bf r} \,\langle \,(\,\nabla \phi({\bf 
r}, t)\,)^2 \,\rangle
\label{eq:energy}
\end{equation}
can be used to determine the dynamic exponent $z$. For vector order 
parameters, $\varepsilon \sim L(t)^{-2}$ at late times,
where $L(t)$ is the length scale beyond which the field of the defect is
screened by other defects.  Figure 1 shows a log-log plot of the energy
density as a function of $t$ for $g = 0,\, 0.5,\, 1,\, 2$ on a $60^3$
lattice. The error bars are smaller than the size of the symbols. Upto
these times ($t=16000$), there is no evidence of finite size effects. The
slight curvature seen in the data (especially for larger $g$) is due to
finite time corrections. The bold line corresponds to $A/(t+t_0)$ where
$A$ and $t_0$ are varied to give the best fit to the data. This shows that
the data gathered over $1.5$-decades gives a $z=2$, {\it independent of}
$g$. 

We next calculate $C(r,t)$ at these late times for different values of
$g$. Figures 2(a)-(b), are scaling plots of $C(r,t)$ versus $r/t^{1/2}$
(see Eq.\ (\ref{eq:scorr})) for $g = 0$ and $g=1$ on a $40^3$ lattice
(finite size effects manifest at $t>12000$). The domain size $L(t)$
extracted from $C(r=L(t), t) = C(0,t)/2$, scales as $t^{1/z}$, where the
exponent is again $z=2$ (within statistical errors) and independent of
$g$. Figure 3 shows that the scaling function $f(x)$ is also {\it
independent of} $g$, for $g = 0,\,0.5,\,1,\,2$. This scaling function is
compared (bold line in Fig.\ 3) with the approximate BPT scaling function for $g=0$ \cite{BPT}, $f(x) =
(3\,\gamma/2\,\pi)\,[\,B(2,1/2)\,]^2 \,F(1/2, 1/2, 5/2\,; \gamma^2)$ where
$\gamma = \exp(-x^2/8)$ and $B$ and $F$ are the Beta and the
hypergeometric functions respectively. 

At larger values of $g$, finite size effects become very prominent. This
can be seen from the form of the late-time $C(r,t)$ for $g=5$ (Fig.\ 4),
simulated on a $50^3$ lattice averaged over $7$ initial configurations.
The correlation function crosses zero at large $r$, dips through a
minimum, and then asymptotically goes to zero (ofcourse $\int C(r,t) >
0$). It is clear from the figure that at these times, $C(r/L(t))$ for
$g=5$ would be qualitatively different from the scaling function of Fig.\
3. However notice that the dip decreases with increasing time. This would
suggest that the dip might disappear at late times \cite{NOTE1}, and that
the resulting scaling function would be identical to Fig.\ 3. In the next
paragraph we will argue that this dip is a preasymptotic feature and
disappears in the scaling limit. This will allow us to conclude that the
scaling function $f(x)$ is indeed independent of $g$. 

At very late times, the order parameter field has totally relaxed with 
respect to defect cores. Preasymptotic configurations typically consist 
of spin wave excitations interspersed between slowly moving defects
separated by a distance $L(t) \gg \xi$, the size of the defect
core. Decomposing $\vec \phi$ into a singular (defect) part $\vec \phi_{sing}$ and a 
smooth (spin wave) part $\vec \phi_{sm}$, we calculate the preasymptotic 
correlation function within a perturbative analysis (see Appendix A for details). 
The computed correlation function exhibits a dip at $r^2/t \sim (1+g^2)/g$, which 
disappears algebraically in time (see Appendix). The amplitude of this dip increases with 
increasing $g$. The dip eventually 
goes away with a relaxation time that scales as ($t_*$ is the time at 
which the $g=0$ correlation function first exhibits scaling),
\begin{equation}
\tau(g) = t_* (1 + g^2)\,.
\label{eq:relax}
\end{equation}
The crossover time $\tau(g)$ is estimated to be (taking $t_*=1000$ for the
$40^3$ system) --- $\tau(g=1)=2000$, $\tau(g=2)=5000$ and
$\tau(g=5)=26,000$. The crossover times for $g \geq 5$ are much larger
than the largest time reached in our simulation ! Figure$\, 5 $ is a
plot of pre-asymptotic  $C(r,t)=C_{sing}+C_{sm}$ at a fixed time where
$C_{sing}$ is given by the BPT form \cite{BPT} and $C_{sing}$ takes
the form derived in Appendix A. Two adjustable parameters related to
the length scale and the amplitude of the spin wave have been tuned to
obtain excellent fits to the numerical data.
Based on these arguments we conclude that the scaling function $f(x)$
is independent of $g$.

We now compute the autocorrelation function $C(0,0,t)$ and extract the
exponent $\lambda$ (see Eq.\ (\ref{eq:sauto})). 
The time $t$ ranges from $4000 - 16000$ for the $60^3$ lattice (averaged
over $50$ initial configurations), well into the scaling regime for the
one-time correlator. Figures 6 are log-log plots of $C(0,0,t)$ versus $t$
for various values of $g = 0,\,0.5,\,1,\,2$. It is difficult to give a precise
value of the decay exponent $\lambda$, since as can be seen from Fig.\ 7,
the ``effective'' $\lambda$ varies by about $3$\% over half a decade.
However it is evident from the bold line in Fig.\ 6, which corresponds to
$A(t+t_0)^{-\lambda/2}$ ($A$ and $t_0$ are varied to give the best fit to
the overall data), that the value of $\lambda$ is independent of $g$. A
fit to each data set for a given $g$, obtains the following values for the
exponent $\lambda$ --- $\lambda(g=0) = 1.526 \pm 0.007$, $\lambda(g=0.5) =
1.521 \pm 0.008$, $\lambda(g=1) = 1.55 \pm 0.01$, and $\lambda(g=2) = 1.55
\pm 0.02$. The numerical values listed above can be compared to the
Mazenko closure estimate of $1.587$ for $g=0$ \cite{BRAY}. These values
obey the Huse-Fisher bound $\lambda > d/2$. 

It is clear that finite size effects set in at later times. As discussed
in \cite{LAMBDA}, finite size effects will be relevant when the spread in
$C(0,0,t)$ (given by $\Delta C(0,0,t) \sim N^{-3/2}$) is of order
$C(0,0,t)$ itself. This will happen when $L^{-\lambda} \sim N^{-3/2}$. The
fact that the numerically computed $\lambda$ increases marginally with
$g$, indicates that finite size effects are more apparent for larger $g$.
This is consistent with our discussion on the effects of finite size on
single-time correlators. 

We end this section with the following assertion based on our careful
numerics. The exponents $z$ and $\lambda$ and the scaling function $f(x)$
are seen to be independent of the torque $g$. This implies that the torque
is irrelevant to the late-time dynamics at $T=0$. In the next section, we
will apply the approximate method of Mazenko to this problem and arrive
at the same conclusion. 

\subsection{Application of the Mazenko Closure Scheme}

Of the variety of approximate schemes devised to evaluate the form of the
scaling function, the closure scheme introduced by Mazenko \cite{MAZENKO}
is amenable to systematic improvement \cite{BRAY,SZ}. We shall use this
closure scheme to determine the scaling form $f(x)$ and show that it is
independent of $g$. The method consists of trading the order parameter
${\vec \phi} ({\bf r},t)$ which is singular at defect sites, for an
everywhere smooth field ${\vec m}({\bf r},t)$, defined by a nonlinear
transformation,
\begin{equation}
{\vec \phi}({\bf r}, t) = {\vec \sigma}\left({\vec m}({\bf r} ,t)\right)\,.
\label{eq:trans}
\end{equation}
At late times, the magnitude of ${\vec \phi}$ saturates 
to its equilibrium value almost everywhere except near the defect cores.
This suggests that the appropriate choice for the nonlinear function ${\vec 
\sigma}$ is an equilibrium defect profile,
\begin{equation}
\frac{1}{2}\nabla_{m}^2 {\vec \sigma}\left({\vec m} ({\bf r} ,t)\right) = 
V^{\prime}\left({\vec \sigma}({\vec m }({\bf r} ,t))\right)\,,
\label{eq:defect}
\end{equation}
where $V^{\prime}(x) \equiv - {\vec x} + ({\vec x} \cdot {\vec x})\,{\vec
x}$. This choice allows for a natural interpretation of ${\vec m}$ (in
the vicinity of a defect) as the position vector from a defect core. The
simplest nontrivial solution of Eq.\ (\ref{eq:defect}) is the hedgehog
configuration,
\begin{equation}
{\vec \sigma}\left({\vec m}({\bf r},t)\right) = \frac{{\vec m}({\bf r}, t 
)} {\vert{\vec m}({\bf r},t) \vert}\,g({\vert{\vec m}\vert})\,,
\label{eq:hedgehog}
\end{equation}
where $g(0)=0$ and $g(\infty)=1$.  Equation (\ref{eq:dyeq2}) can be
used to derive an equation for the correlation function $C(12) \equiv
\langle {\vec \phi}({\bf r}_1, t_1) \cdot {\vec \phi}({\bf r}_2,
t_2)\rangle$. Substituting for $\phi$ (Eqs.\ (\ref{eq:trans}) and 
(\ref{eq:hedgehog})) in the right hand side of the
resulting equation, we get
\begin{eqnarray}
{\partial_t}{C(12)}   &  =  &
                            \nabla_1^2 C(12) + \langle\, {\vec \sigma}
                            ({\vec m}(2)) \cdot V'({\vec \sigma}
                            ({\vec m}(1)))\,\rangle \nonumber \\
                      &     &
                            +\, g \,\langle\, {\vec \sigma}({\vec m}(2)) 
                           \cdot {\vec \sigma}({\vec m}(1)) \times 
                           \nabla^2 {\vec \sigma}({\vec m}(2))\,\rangle\,\,. 
\label{eq:langm}
\end{eqnarray}

So far no approximation has been made, but further progress seems
impossible without one.  Now along with Mazenko, we make the assumption
that each component of ${\vec m}({\bf r},t)$ is an independent gaussian
field with zero mean at all times. This implies that the joint probability
distribution $P(12) \equiv P({\vec m}(1), {\vec m}(2))$ is a product of
separate distributions for each component and is given by  \cite{BRAY},

\begin{equation}
\prod_{\alpha} {\cal N} \exp
      \left\{\, - \frac {1} {2\left(1-\gamma^{2}\right)} \left(
     \frac{m_{\alpha}^2(1)}{S_0(1)}+\frac{m_{\alpha}^2(2)}{S_0(2)}\\
        - \frac{2 \gamma m_{\alpha}(1)m_{\alpha}(2)} 
        {\sqrt{S_0(1)S_0(2)}} \right) \right\}\,,
\label{eq:mazprob} 
\end{equation}
where $$ {\cal N} = \frac{1}{2\pi\sqrt{(1-\gamma^2)S_0(1)S_0(2)}} $$
and \[ \gamma \equiv \gamma(12) = \frac {C_0(12)}{\sqrt{S_0(1)S_0(2)}}.\]
The joint distribution has been written in terms of the second moments
$S_0(1) = \langle m_{\alpha}(1)^2 \rangle$, $S_0(1) = \langle 
m_{\alpha}(1)^2 \rangle$ and $C_0(12) = \langle m_{\alpha}(1)
m_{\alpha}(2) \rangle$.

With this assumption, the right hand side of Eq.\ (\ref{eq:langm}) 
simplifies to,
\begin{eqnarray}
\frac{\partial C(12)}{\partial t_1} & = & \nabla^2 C(12)
                   + \frac{\gamma}{2S_0(1)}\frac{\partial
                     C(12)}{\partial \gamma } \nonumber \\
             &  &  +\,g\, \langle \,{\vec \sigma}({\vec m}(2)) 
                     \cdot {\vec \sigma}({\vec m}(1))
	 	 \times \nabla^2 {\vec \sigma}({\vec m}(2))\,\rangle\,.
\label{eq:mazcorr}
\end{eqnarray}
It is clear that since $P({\vec m}(1), {\vec m}(2)) = P({-\vec m}(1),
{-\vec m}(2))$, the last term, which is odd in ${\vec m}$, drops out. The
resulting equation is identical to the purely dissipative one, showing
that the torque is irrelevant at late times. This conclusion, consistent
with our earlier numerics, is a direct consequence of the Mazenko closure
approximation.

\subsection{Justification of the Mazenko Approximation for vector order 
parameters}
In this section we will subject the above stated assumptions of the Mazenko 
closure scheme to systematic numerical tests. We will do this by numerically
solving the Langevin equation Eq.\ (\ref{eq:dyeq2}) by the method 
outlined in Section III A. Knowing ${\vec \phi}({\bf r},t)$, one can compute 
${\vec m}({\bf r},t)$ by inverting Eq.\ (\ref{eq:hedgehog}). This is  
facilitated by choosing
\begin{equation}
g({\vert{\vec m}\vert}) = \frac{\vert{\vec m}\vert}{\sqrt{1+\vec m \cdot 
\vec m}} 
\label{eq:formg}
\end{equation}
which is consistent with the boundary conditions for $g(x)$ mentioned in the
previous section. The resulting ansatz for $\vec \sigma$,
\begin{equation}
\vec \sigma(\vec m({\bf r},t))= \frac{\vec m}{\sqrt{1+\vec m \cdot \vec 
m}}\,,
\label{eq:mazsig} 
\end{equation} 
can be easily inverted. We calculate both the single point probability
distribution $P({\vec m}({\bf r},t))$ and the joint probability
distribution $P(12)$ in the scaling regime and compare with the Mazenko 
assumption.
\newpage
In what follows all probability distributions have been computed  
on a $45^3$ lattice and averaged over $100$ initial configurations.
We shall display the distributions for $g = 0$ and $g=1$. We have collected
data in the scaling regime from $t = 2000$ to $t = 15000$, after which
finite size effects set in.

Figure 8 and 9 are scaling plots of $P(m_1({\bf r},t))$ at $g=0$ and $g=1$
respectively. In accordance with the Mazenko assumption, the scaling variable has
been taken to be $m_1/\sqrt{S_0(t)}$, where $S_0(t) = \langle m_{1}(r,t)^2
\rangle$. A plot of $S_0(t)$ (for both $g=0$ and $1$) versus $t$ (Fig.\ 10), shows
a linear growth over a decade, consistent with a $z=2$. The scaled distribution
$P(m_1)$ is also seen to be the same for $g=0$ and $g=1$ (Fig. 11), suggesting
that it is independent of $g$. 

Though the distribution grossly resembles a gaussian at late times, a closer
inspection shows systematic deviations at small values of $m_{1}$
(Fig.\, 11). 
The distribution seems to be flatter than a gaussian when $m_1 \approx 0$.
This is clearly visible in a plot of $ -\ln(\,-\ln\left[P(m^2_1/S_0)
\right]\,)$ versus $\ln(m^2_{1}/S_0)$, which shows that
 distribution deviates from a gaussian for small $m_{1}$ (Fig.\, 12). 
These findings are consistent with a similar
analysis done on a scalar order parameter \cite{CHUCK}.

We now study the two point distribution $P({\vec m}(1),{\vec m}(2))$.
Equation (\ref{eq:mazprob}) implies that in the variables 
\[{\vec m}_{\pm}(12) =\frac{1}{2}\left(\frac{{\vec m}
(1)}{\sqrt{S_{0}(1)}} \pm \frac{{\vec m}
(2)}{\sqrt{S_{0}(2)}} \right)(S_{0}(1)S_{0}(2))^{\frac{1}{4}}\:,  \]
the distribution $P(12)=P({\vec m}_{+}(12))P({\vec m}_{-}(12))$, where

\begin{eqnarray}
 P({\vec m}_{\pm}(12)) & = & \prod_{\alpha}{\sqrt {\cal N }}\exp \left\{ \:
-\frac{m^2_{\alpha \pm }}{(1\pm\gamma)\sqrt{S_{0}(1)S_{0}(2)}} \right \}
\end{eqnarray} 

We compute $ P(m_{\alpha+}(12), m_{\alpha -}(12))$
at $t_1 = t_2 = t $ in the scaling regime. A plot of $
P(m_{1+},m_{1-}) $ (Fig.\ 13) for $ \vert{\bf r}_1-{\bf r}_2\vert = 4\sqrt{3}
$ looks like a product of two gaussian distributions. To check this we
compute the difference
\begin{eqnarray}
\Delta (m_{\alpha+}(12), m_{\alpha -}(12)) & = & P( m_{\alpha+}(12),
m_{\alpha -}(12)) \nonumber \\ 
	&    & - P(m_{\alpha +}(12))P( m_{\alpha-}(12))\,, \nonumber
\end{eqnarray} 
which should be zero everywhere if the Mazenko approximation were to hold.

Figure 14 shows a surface plot of the difference $\Delta(m_{1+}(12),
m_{1-}(12))$, magnified $10^5$ times. It is clear that $
\Delta(m_{1+}(12), m_{1-}(12))$ is zero everywhere except in the region
close to the origin, where the maximum deviation from $0$ is around
$10^{-5} \:$. The situation is similar for $ g = 1 $. We now compute $
P(m_{1+}(12))$ and $ P( m_{1-}(12))$, and find that the scaled
distributions are the same for $ g = 0 $ and $ g = 1 $. Figures $ 15 $ and
$17$ are scaling plots of $ P(m_{1+}) $ and $ P(m_{1-}) $ respectively and
indicate that the scaling function is independent of $g$. Moreover the
distribution looks like a gaussian, in accordance with the Mazenko theory.
Figures 16 and 18 are plots of $ -\ln(-\ln[P(m_{1\pm}^2(12)/\langle
m_{1\pm}^2(12) \rangle)]) $ versus $ \ln(m_{1\pm}^2(12)/\langle
m_{1\pm}^2(12) \rangle) $. The deviation from a straight line when $
m_{1\pm} \approx 0$, indicates that the distributions differ slightly from
a gaussian. Note that the data for small $m_{1\pm}$ in Figs.\ $16, 18$ do
not quite scale and so it is likely that in the computation of the
joint-probability distribution, we have not yet reached the scaling
regime.

In conclusion, we have shown, by computing the single and the two -point
probability distributions, that the Mazenko closure scheme is a fairly
good approximation at late times with possible deviations when $
m_{\alpha} \approx 0 $. This study justifies the use of the Mazenko
approximation in the evaluation of the correlation functions and hence the
results of the previous section. This constitutes the first serious
numerical check on the assumptions of the Mazenko scheme for vector order
parameters. 

\section{ Phase Ordering Dynamics\,: Quenches to $T = T_c$}

We now quench from the high temperature paramagnetic phase to
the critical point $T_c$, and ask whether the spin precession
changes the late time dynamical behaviour. More precisely, we
would like to know whether the torque term $g$ is relevant at the
Wilson-Fisher fixed point corresponding to the pure $g=0$ Heisenberg
model with nonconserved dynamics. 

This can be done by power counting \cite{HKJ} to 
${\cal O}(\epsilon^2)$ at 
the Wilson-Fisher fixed point $r^{*}= -(5/22)\Lambda^2 \epsilon$,
$u^{*}= 8 \pi^2 \epsilon/11$ ($\Lambda$ is the ultraviolet cutoff). 
Dimensional analysis provides the scaling
dimension $[g] = d/2 + 1 - z + \eta/2$. At the Wilson-Fisher fixed point 
$z=2+c\eta$, where $c=6 \ln (4/3) -1$, and $\eta = (5/242) \epsilon^2$.
This implies that above $d=2$, the torque term is irrelevant. In $d=1$,
the torque is relevant, as can be seen by an explicit solution of the 
Langevin equation Eq.\ (\ref{eq:dyeq2}) in one spatial dimension.

\section{Conclusions}

We have investigated the effect of reversible Possion bracket terms in a
generalized Langevin equation on the phase ordering dynamics at late
times. The dynamics of such systems can be classified by the structure of the
Poisson brackets. In this Part I of two parts, we have made a detailed
study of the nonconserved dynamics of an order parameter whose
components obey a Lie Algebra. A common example is the nonconserved
dynamics of a Heisenberg magnet. The dynamics consists of two parts
--- an irreversible 
dissipation into a heat bath and a reversible precession induced by a
torque due to the local molecular field. For quenches to zero temperature,
we have shown, both numerically (Langevin simulation) and analytically
(approximate closure scheme due to Mazenko), that the torque is irrelevant
at late times. The Mazenko closure was subject to critical numerical
tests and was shown to perform well at late times (apart from small
deviations). For
quenches to $T_c$, we show to
${\cal O}(\epsilon^2)$, that the torque is irrelevant at the Wilson-Fisher
fixed point. 

In Part II of this series we investigate the effect of the reversible
torque on the conserved dynamics of the Heisenberg magnet in three
dimensions. We will show that the torque is relevant both for quenches to
$T=0$ and $T=T_c$. 

\bigskip
\begin{appendix}
\section{}

In this Appendix we study the effect of spin wave excitations on the time
dependent preasymptotic equal-time correlation function. We would like to show that inclusion
of such excitations in the correlation function $C(r,t)$, leads to a
dip at preasymptotic
times when $g \ne 0$, which eventually relaxes.  As suggested in Section III A,
at very late times, the order parameter field has totally relaxed with respect to
defect cores.  Preasymptotic configurations typically consist of spin wave
excitations interspersed between slowly moving defect cores. These defect cores
are separated by a typical distance $L(t) \gg \xi $, the size of the defect core. In
general one can decompose $\vec \phi({\bf r}, t) = \vec \phi_{sing}({\bf r}, t) +
\vec \phi_{sm}({\bf r}, t)$, where the singular part $\vec \phi_{sing}$ 
parametrizes defect configurations while the smooth part $\vec \phi_{sm}$ is a 
linear combination of spin waves of wave-vector ${\bf k}$, $\vec \phi_{sm}({\bf 
r}, t) = V^{-1/2} \sum \vec \phi^{sm}_{\bf k}(t) e^{i {\bf k}\cdot{\bf r}}$.
The preasymptotic correlation function will thus have three contributions ---
$C_{sing} \equiv \langle \vec \phi_{sing}(0, t) \cdot \vec \phi_{sing}
({\bf r}, t) \rangle$, $C_{sm} \equiv \langle \vec \phi_{sm}(0, t) \cdot \vec 
\phi_{sm}({\bf r}, t) \rangle$ and the scattering of spin waves from slowly 
moving defects $C_{scat} \equiv \langle \vec \phi_{sm}(0, t) \cdot \vec 
\phi_{sing}({\bf r}, t) \rangle$. At late times of course $\vec \phi^{sm}_{\bf k}
(t) \to 0$, and $\vec {\phi}_{sing}$ can be traded off for the auxilliary field
$\vec m$ within the Mazenko approach. Thus the $C_{sing}(r, t)$ part of the 
correlation function is given by the solution of Eq.\
(\ref{eq:mazcorr}) or the BPT form \cite{BPT}.

The smooth part of the correlation function $C_{sm}(r, t)$ can be estimated
from a perturbative calculation, wherein the defects separated by a
distance $L \gg \xi$ are taken to be static (justified post priori). We
shall see that the dip is a result of this smooth part. Confining our
attention to a single domain of size $L$, we can split the smooth $\vec
\phi_{sm}$ into transverse and longitudinal components about the
well-defined broken symmetry axis taken to be along $\alpha = 3$. Thus
$\phi^{sm}_{\alpha} ({\bf r}, t) = \phi_{eq} \,\delta_{\alpha\,3} +
u_{\alpha}({\bf r},t)$, where the equilibrium magnetisation $\phi_{eq} =
1$. 

Consider an initial smooth localised pulse in the interior of this domain
of the form $u_{1} ({\bf r }, 0) = u_{2} ({\bf r }, 0) = \frac {u(0)}
{(2\sigma)^3\pi^{3/2}} e^{-r^2/4\sigma^2}$ and $u_{3} ({\bf r}, 0) = \frac
{u_{3}(0)} {(2\omega)^3\pi^{3/2}} e^{-r^2/4\omega^2}$, where the widths
$\sigma,\,\omega \ll L$ and $u(0),\, u_3(0) \ll 1$. The equation for
$u_{\alpha}$ can be read out from Eq.\ (\ref{eq:dyeq2}),

\begin{eqnarray}
\frac{\partial u_{\alpha}}{\partial t } & = & \nabla^2 u_{\alpha}
-g\epsilon_{\alpha \beta 3 }\nabla^2u_{\beta}-2u_{3}\delta_{\alpha 3 }
\nonumber \\ 
&  & -(u_{\beta}u_{\beta})\delta_{\alpha
3}-2(u_{\alpha}u_{\beta})\delta_{\beta3} \nonumber \\ 
& & +g\epsilon_{\alpha\beta\gamma}u_{\beta}\nabla^2u_{\gamma} 
-u_{\beta}u_{\beta}u_{\alpha}. 
\label{eq:spinwav}
\end{eqnarray}

To solve this equation perturbatively, we multiply the nonlinear terms in
Eq.\ (\ref{eq:spinwav}) by an arbitrary real parameter $\epsilon \,(\leq
1)$ and express $u_{\alpha}({\bf r}, t)$ as $\displaystyle
\sum_{n=0}^{n=\infty}\epsilon^{n}u^{(n)}_{\alpha}({\bf r }, t)$. The
initial conditions for $u^{(n)}_{\alpha}({\bf r }, 0)$ follow from
$u_{\alpha}(r, 0)$. We are interested in solutions that decay as $t \to
\infty$. Convergence of the perturbation series at $\epsilon = 1$ is
guaranteed by the smallness of the initial deviation and because
higher-order terms in the expansion decay faster. To ${\cal
O}(\epsilon^0)$, the spin waves do not interact and

\begin{eqnarray}
\frac{\partial u^{(0)}_{\alpha}}{\partial t } & = & \nabla^2 u^{(0)}_{\alpha}
-g\epsilon_{\alpha \beta 3 }\nabla^2u^{(0)}_{\beta}-2u^{(0)}_{3}
\delta_{\alpha 3 }\,.
\end{eqnarray}	

The equations for $u^{(0)}_{1}$ and $u^{(0)}_{2}$ decouple in the
variables $u^{(0)}_{\pm} = (u^{(0)}_{1} \pm iu^{(0)}_{2})/2$, giving rise
two precessing goldstone modes in the transverse direction and an 
exponentially decaying mode in the longitudinal direction.
Thus,
\begin{eqnarray}
 u^{(0)}_{\pm}({\bf r },t) & = &
\frac{(1 \pm i)u(0)}{16(\pi t(1\pm
ig))^{3/2}}\exp\left(-\frac{r^2}{ 4(1\pm ig)t}\right) \nonumber
\end{eqnarray}
and
\begin{eqnarray}
u^{(0)}_{3}({\bf r },t) & = &
\frac{u_{3}(0)}{8(\pi t)^{3/2}}\exp\left(-\frac{r^2}{4t}-2t\right) \nonumber \\
\label{eq:zero}
\end{eqnarray}
are the asymptotic solutions to ${\cal O}(\epsilon^0)$.

To ${\cal O}(\epsilon)$, the dynamical equations in the transverse variables
are given by,

\begin{eqnarray}
\frac{\partial u^{(1)}_{\pm}}{\partial t } &  =  & (1\pm
 ig) \nabla^2 u^{(1)}_{\pm} - 2\,u^{(0)}_{3}\,u^{(0)}_{\pm}
 \pm ig(u^{(0)}_{3}\,\nabla^2 u^{(0)}_{\pm} \nonumber \\
&  & -\,u^{(0)}_{\pm}\nabla^2u^{(0)}_{3})-
\,(\,4\,u^{(0)}_{+}u^{(0)}_{-}+(u_3^{(0)})^2\,)\,u_{\pm}^{(0)}
\label{eq:peq1}
\end{eqnarray}

The last two terms are subdominant in $1/t$ and $u(0), u_3(0)$ 
respectively, and so the transverse correlator to ${\cal O}(\epsilon)$ is 
given by,
 
\begin{eqnarray}
C_{sm}^{\perp}({\bf r },t) &  =  & \frac{1}{2} \langle u_{+}({\bf x 
},t)u_{-}({\bf 
x+r},t) + u_{+}({\bf x+r },t)u_{-}({\bf x},t)\rangle \nonumber \\
                   & \sim & \frac{A_{1}}{t^{3/2}}\exp \left(
-\frac{r^2}{8 t } \right)+\frac{A_{2}e^{-2t}}{t^{3}}\exp \left(
-\frac{r^2}{2t(3+g^2)} 
\right)\nonumber  \\
                        &    &
 \times\left\{\cos \left(
\frac{gr^2}{4t(3+g^2)} + \pi/4 \right)\right\}\,,
\label{eq:acperp}
\end{eqnarray}
where $A_1 \sim O(u(0)^2)$ and $A_2 \sim O(u(0)^2\,u_3(0))$
are constants depending on $g$ and initial conditions.
The cosine term in the above expression results in the observed dip of the
total correlation function. The magnitude of the dip increases
with increasing $g$. 

The dynamical equation for the longitudinal component to 
${\cal O}(\epsilon)$ is likewise given by,

\begin{eqnarray}
 \frac{\partial u^{(1)}_{3}}{\partial t } & = &
 (\nabla^2-2)\,u^{(1)}_{3} - (4u^{(0)}_{+}\,u^{(0)}_{-}
 + \,3\,(u^{(0)}_{3})^2) \nonumber \\
  &  & + 2\,ig\,(u^{(0)}_{+}\,\nabla^2u^{(0)}_{-}\,-
 u^{(0)}_{-}\,\nabla^2u^{(0)}_{+}) \nonumber \\
  &  &-\,(\,4u^{(0)}_{+}u^{(0)}_{-}+(u_3^{(0)})^2\,)\,u_3^{(0)}
\end{eqnarray}

As before, the terms proportional to $(u^{(0)}_{3})^2$, the gradient terms 
and the cubic term are subdominant, and so the decay of the longitudinal 
correlation function $C_{sm}^{\parallel}$ is given by,

\begin{eqnarray}
C_{sm}^{\parallel}({\bf r },t) & =  & \langle \phi_{3}({\bf x},t)
 \phi_{3}({\bf x+r},t) \rangle \nonumber \\
 & \sim & \frac{B_1e^{-4t}}{t^{3/2}}\exp\left(-\frac{r^2}{8t}\right) -
 \frac{B_2e^{-2t}}{t^3}\exp
 \left(-\frac{r^2}{2t(3+g^2)} \right) \nonumber \\
   &  & +\frac{B_3}{t^{9/2}}\exp\left(-\frac{r^2}{4t(1+g^2)}\right)\,,
\label{eq:acpar}
\end{eqnarray}
where $B_1\sim O(u_3(0)^2, u(0)^2)$, $B_2 \sim O(u(0)^2\,u_3(0), u(0)^3)$ 
and $ B_3 \sim O(u(0)^4)$ are funtions of $g$ and initial conditions.

Note that $C_{sm}^{\perp}$ evolves with a width that scales as 
$t^{1/2}$ and whose amplitude decreases as $t^{-3/2}$. The longitudinal   
$C_{sm}^{\parallel}$ decays exponentially fast. This decay is consistent with our
earlier assertion that the defects separated by a distance $L(t)$ hardly
move over time scales corresponding to spin wave relaxation. 

The cross correlator $C_{scat}$ coming from the scattering of spin waves by moving
defects, can be calculated by treating $\phi_{sm}$ as ``slaved'' to $\phi_{sing}$.
As the defects move they excite spin waves which decay in a time scale smaller
than the time taken by the defects to move any further. The dominant contribution
to $C_{scat}$ comes from the product of Eq.\ (\ref{eq:zero}) and $\phi_{sing}$. It
is easy to see that this term leads to the same cosine dip as in Eq.\
(\ref{eq:acperp}) but with an amplitude that decays algebraically in time. This is
the source of the slow decay of the dip. 
\end{appendix}

\newpage

\begin{figure}
\centerline{\psfig{figure=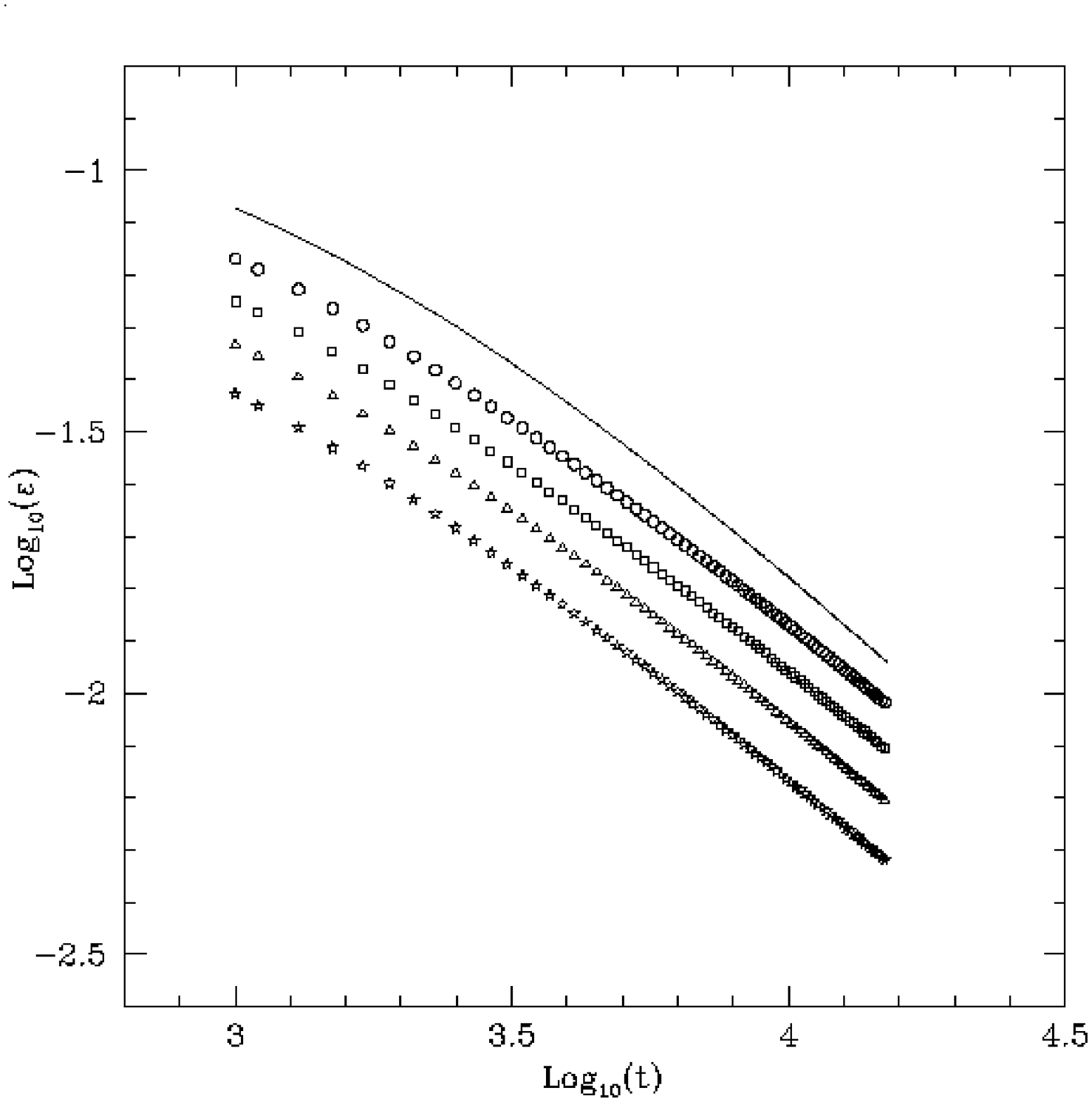,width=7.5cm,height=9.0cm}}
\end{figure}
Fig. 1 log-log plot of the energy density $\varepsilon$
versus $t$ for various values of $g,\: g=0 (\circ) ,\: g=0.5
 (\Box),\: g=1 (\triangle),\: g=2 (\star ) $. The straight line is a fit (see
text).

\begin{figure}
\centerline{\psfig{figure=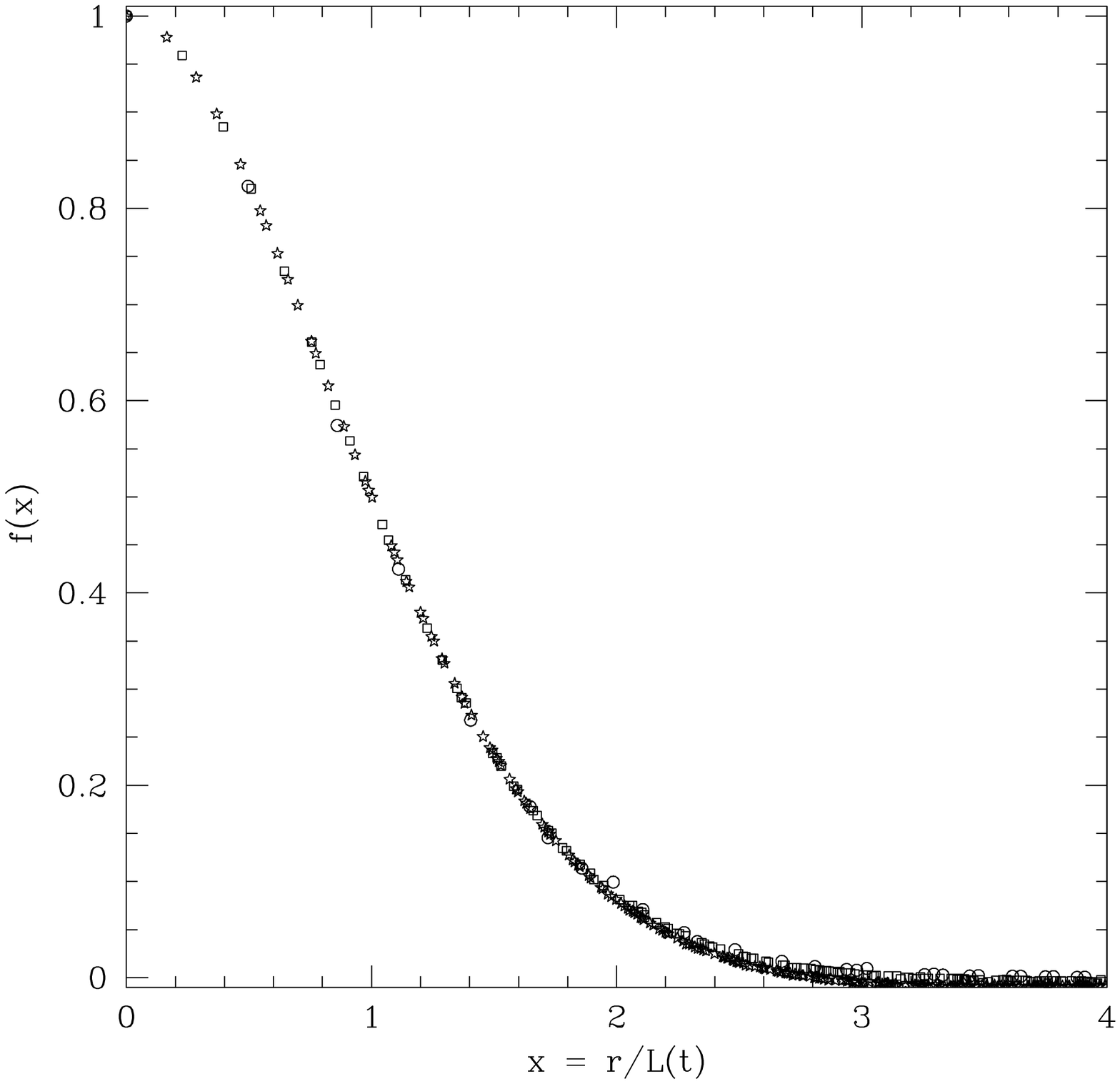,width=8.0cm,height=8.0cm}}
\end{figure}
Fig. 2(a). The scaling function $f(x)$ versus
$x\equiv r/t^{1/2}$  for $g=0$. 
Data taken at $t=1000\,(\circ)$, $t=5000\,(\triangle)$ and 
$t=10000\,(\star)$. Error bars are smaller than the size of symbols.

\begin{figure}
\centerline{\psfig{figure=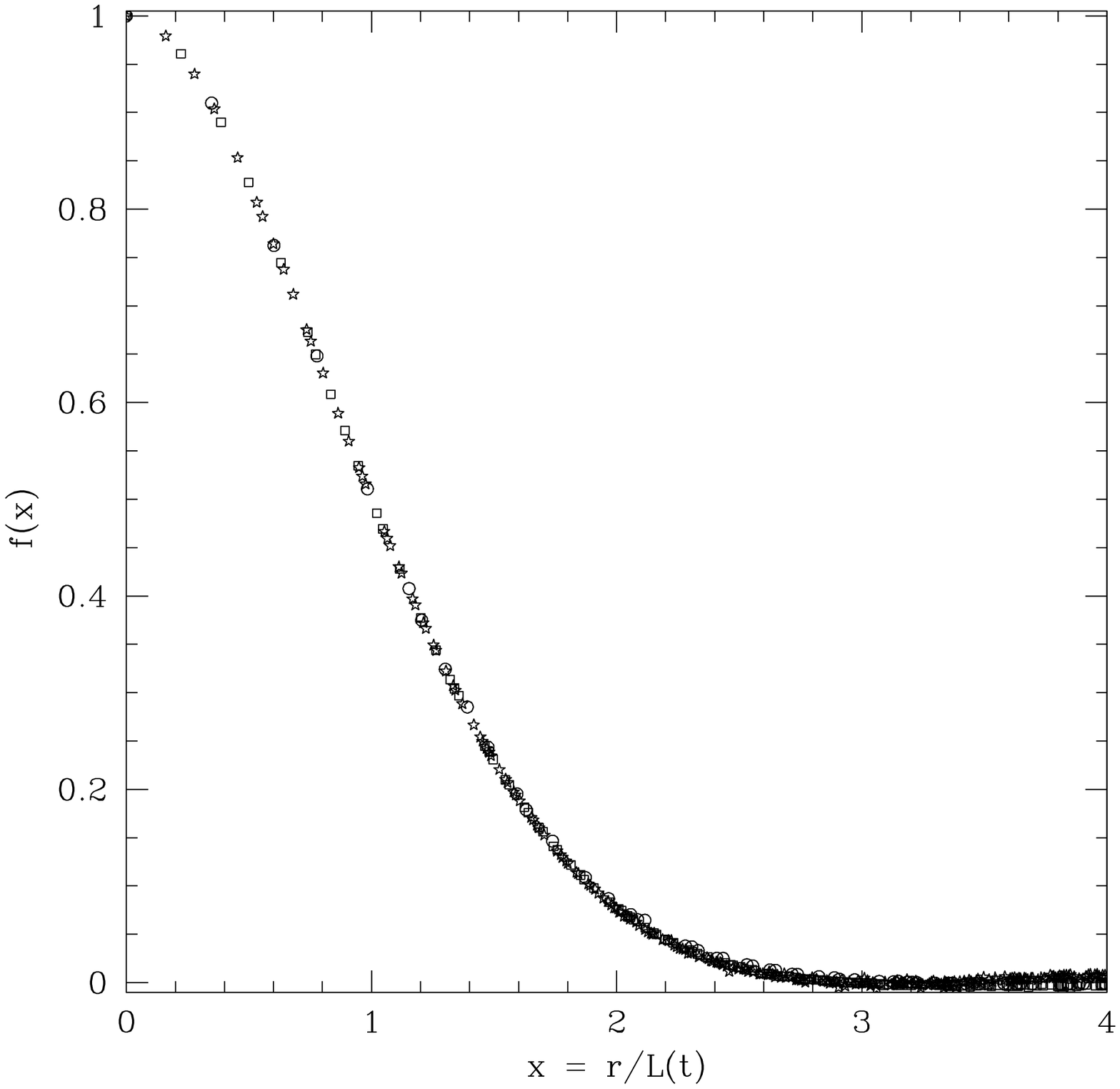,width=8.0cm,height=8.0cm}}
\end{figure}
Fig. 2(b). $f(x)$ versus $x \equiv r/t^{1/2}$ for $g=1$. 
Data taken at $t=2000\,(\circ)$, $t=5000\,(\triangle)$ and 
$t=10000\,(\star)$. Error bars as in Fig.\ 2(a).

\begin{figure}
\centerline{\psfig{figure=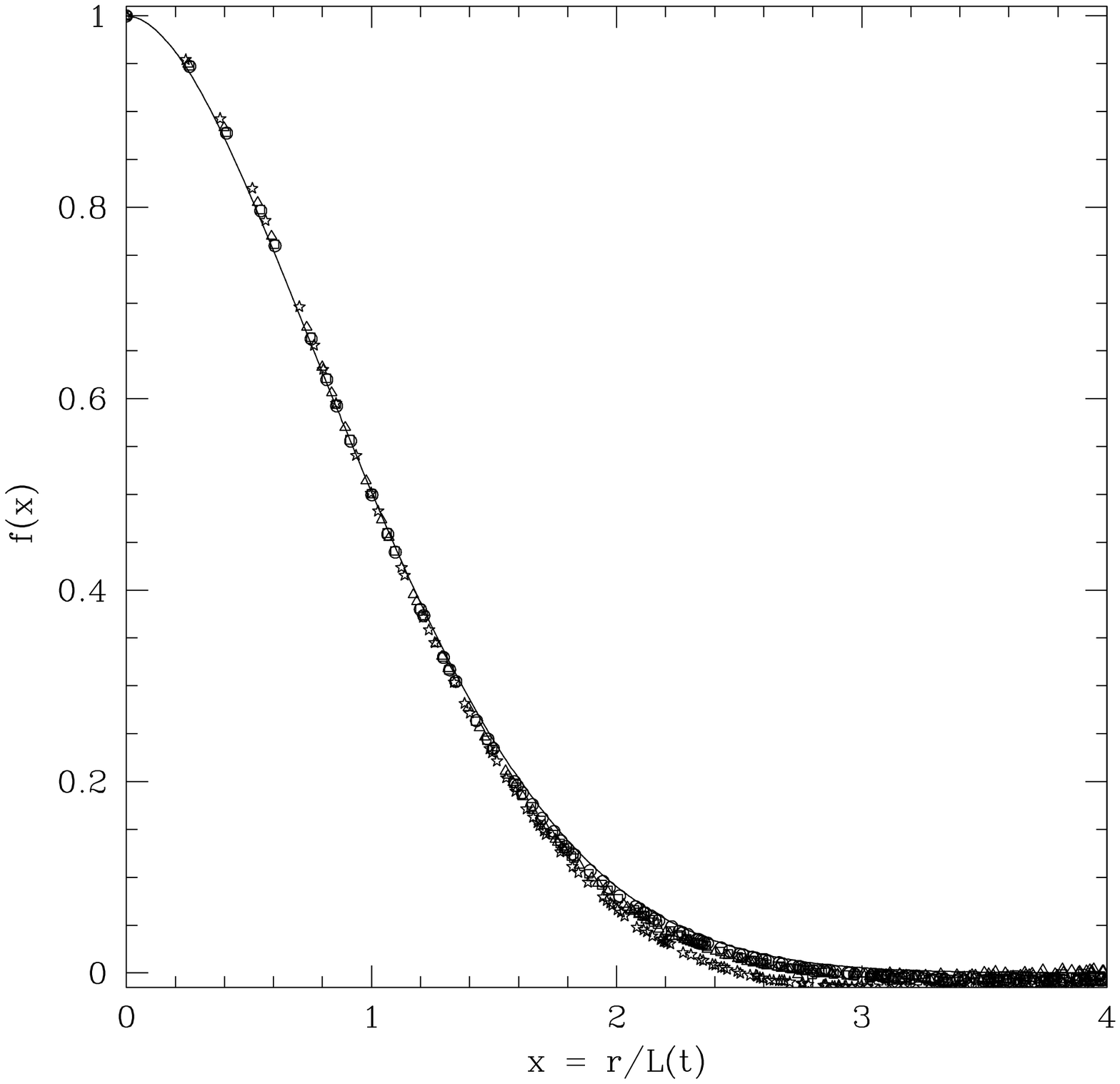,width=8.0cm,height=8.0cm}}
\end{figure}
Fig. 3. $f(x)$ versus $x\equiv r/t^{1/2}$ for different values of $g$ ($g =
0\,(\circ),\:0.5\,(\Box),\:1\,(\triangle),\:2\,(\star)$).  The data for
$g=2$ was obtained from a simulation on a $50^3$ lattice averaged over $7$
configurations. The continuous curve is the approximate analytical form
defined in the text \cite{BPT}. 

\begin{figure}
\centerline{\psfig{figure=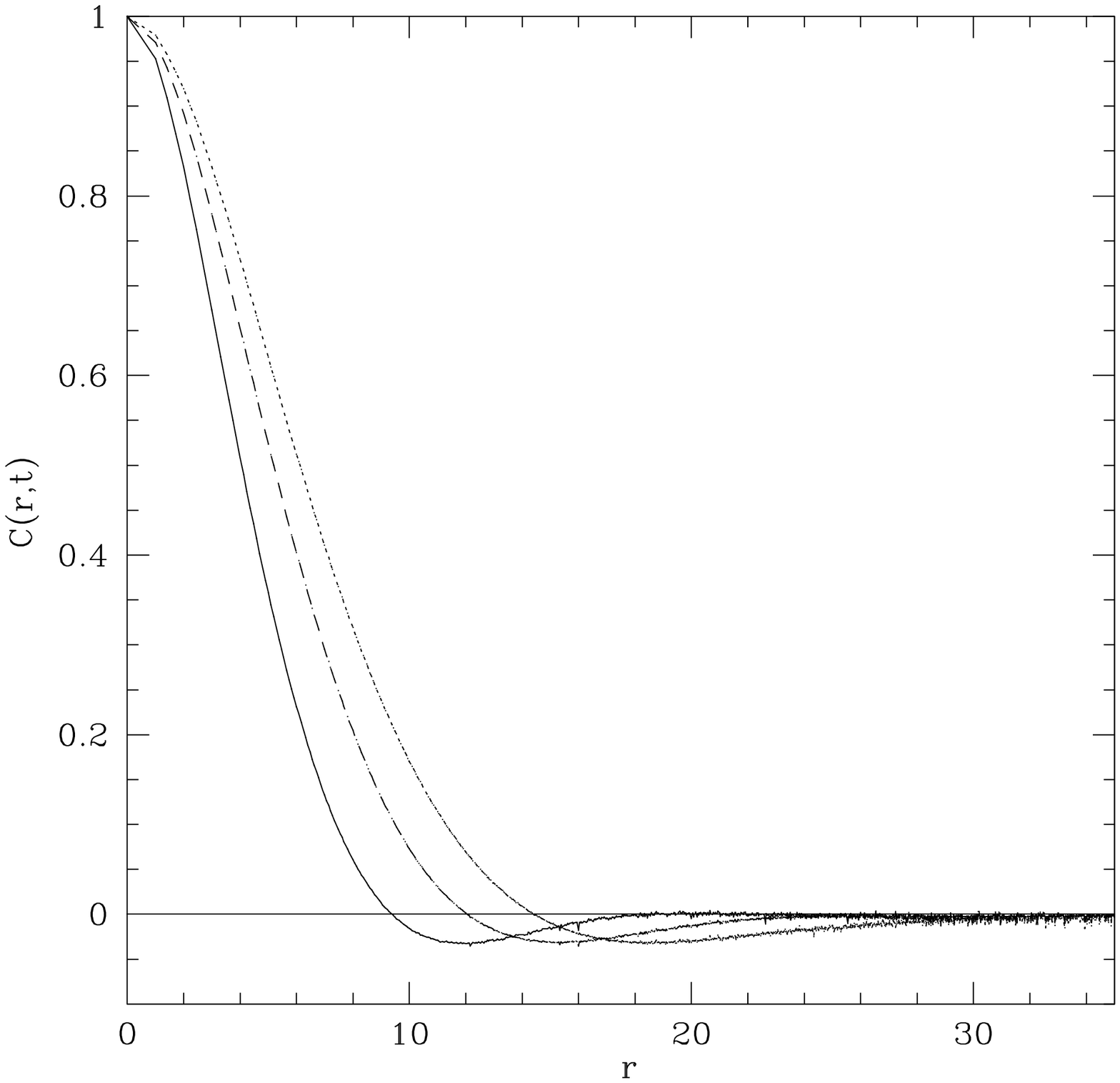,width=8.0cm,height=8.0cm}}
\end{figure}
Fig. 4. The
correlation function $C(r,t)$ versus $r$ for $g=5$ at various times
($\:t=3000\:$(\rule[0.6mm]{5mm}{.1mm}),$\:t=5000\:(-\cdot-\cdot), \:t=7000\:( \cdots)$.  Note
that the dip gets smaller as time progresses. Error bars as in Fig.\
2(a).

\begin{figure}
\centerline{\psfig{figure=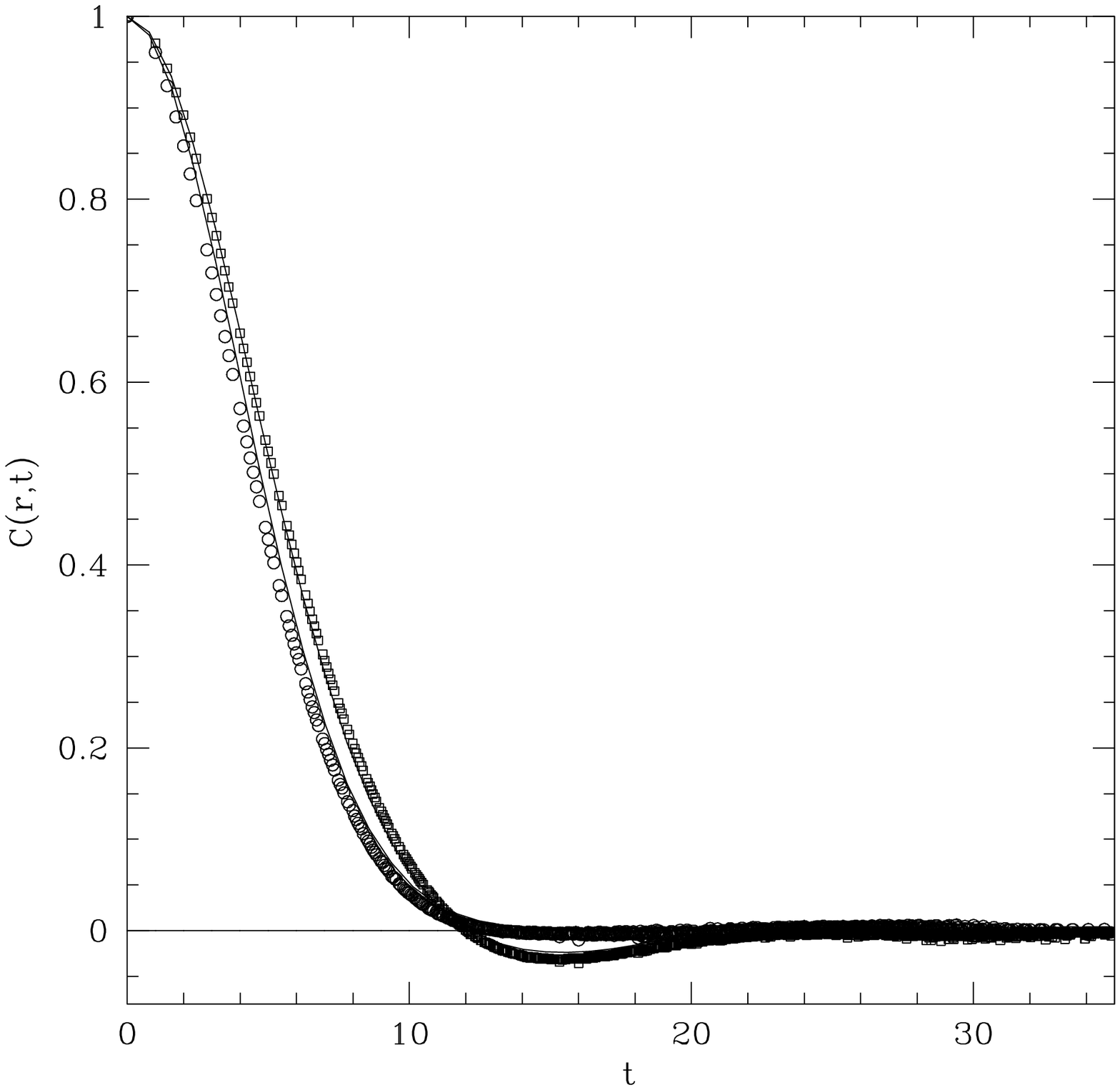,width=8.0cm,height=8.0cm}}
\end{figure}
Fig. 5. Preasymptotic $C(r,t) $ for $ g=1$ and $5$ (
bold lines ) calculated in Appendix A. Note the excellent fit to data for
$ g=1\,(\circ),\,g=5\,(\Box) $

\begin{figure}
\centerline{\psfig{figure=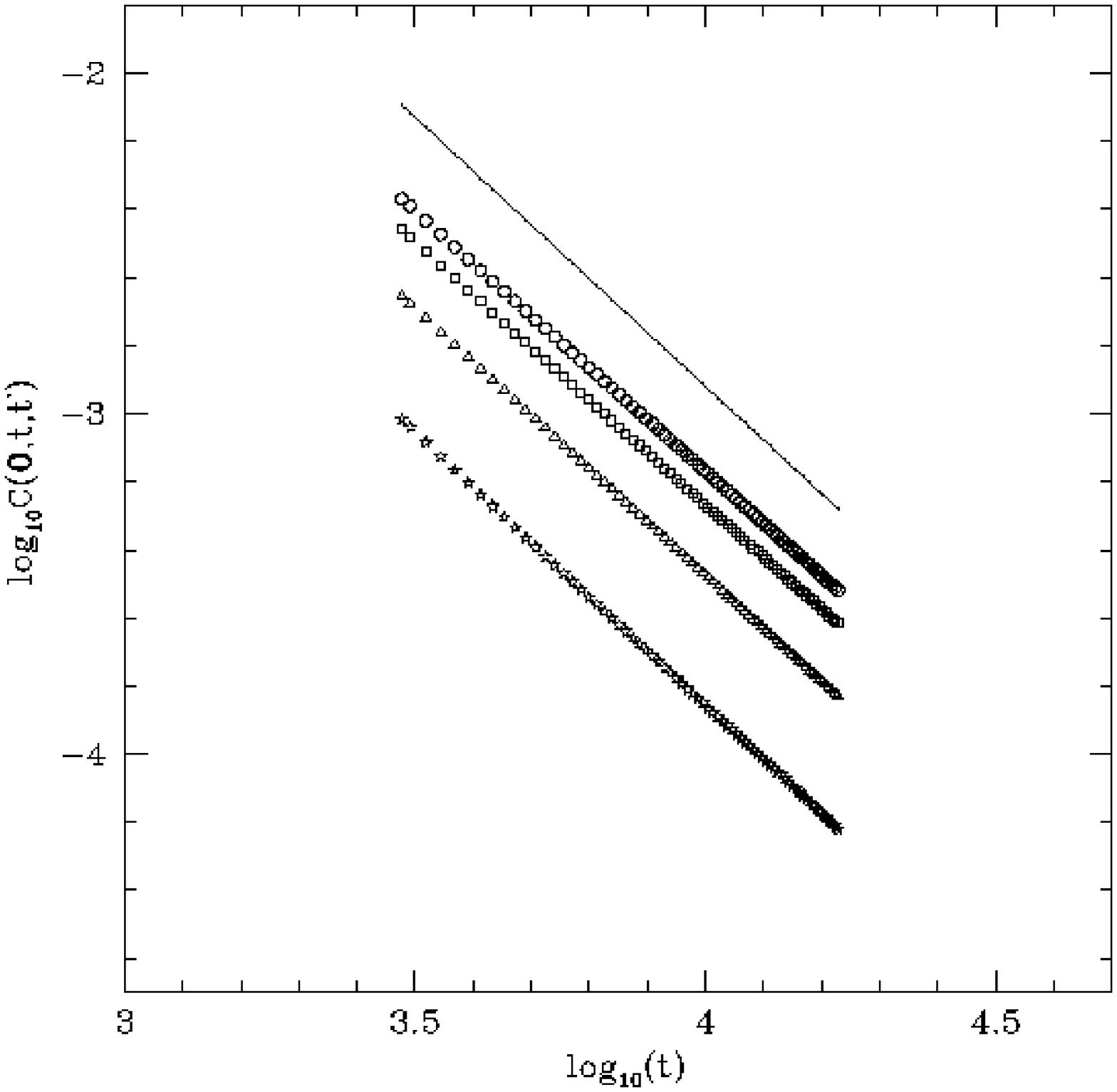,width=7.5cm,height=9.0cm}}
\end{figure}
Fig. 6. Log-Log plot of the autocorrelation function 
$C(0,0,t)$ versus $t$ for $g=0\,(\circ),\: 0.5\,(\Box),\: 1\,(\triangle),\:
2\,(\star)$. The straight line is a fit $A(t+t_0)^{-\lambda/2}$,
where $\lambda$ has been chosen to be the Mazenko value $-1.587$.

\begin{figure}
\centerline{\psfig{figure=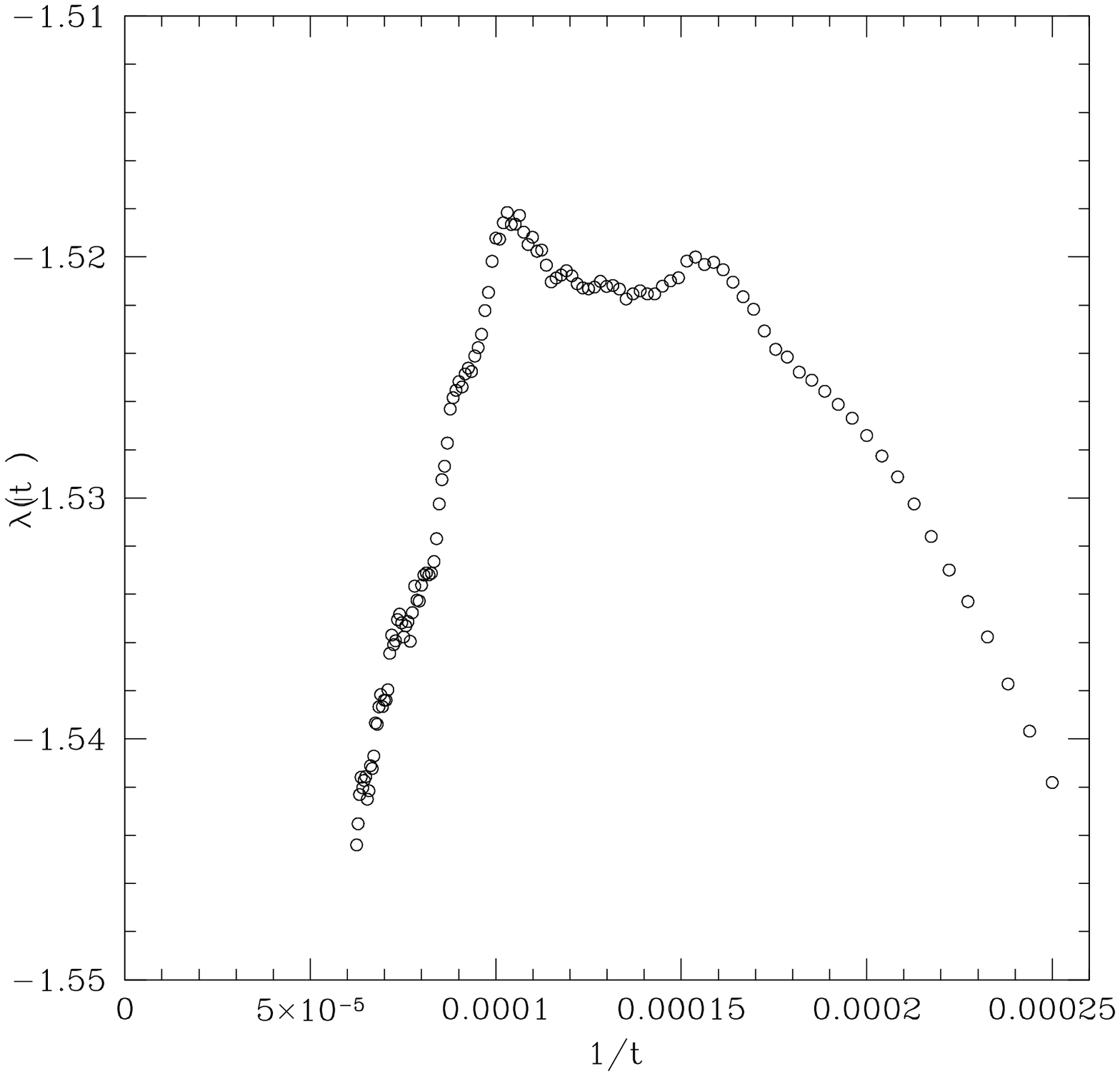,width=8.0cm,height=8.0cm}}
\end{figure}
Fig. 7. Effective $\lambda$ as a function of $1/t$ for 
$g=0$.

\begin{figure}
\centerline{\psfig{figure=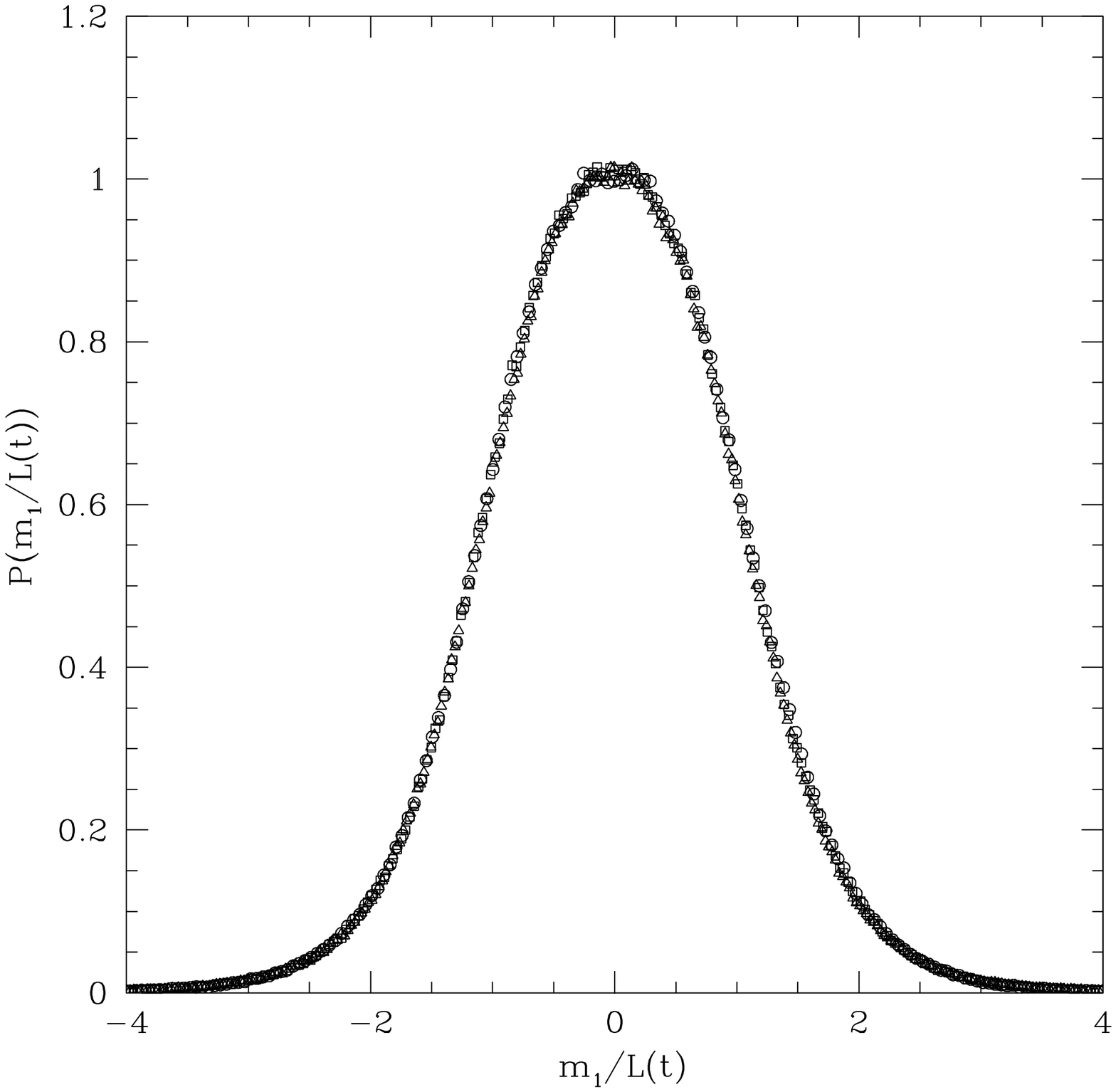,width=8.0cm,height=8.0cm}}
\end{figure}
Fig. 8. Single point distribution of $P(m_1)$ for $g=0$ at different
times $\: t=5000\,(\circ), \:t=10000\,(\Box), \:t=15000\,(\triangle)\:$. 
The distribution scales in the variable $m_1/L$,
where $L(t) = \sqrt{S_0(t)}$.

\begin{figure}
\centerline{\psfig{figure=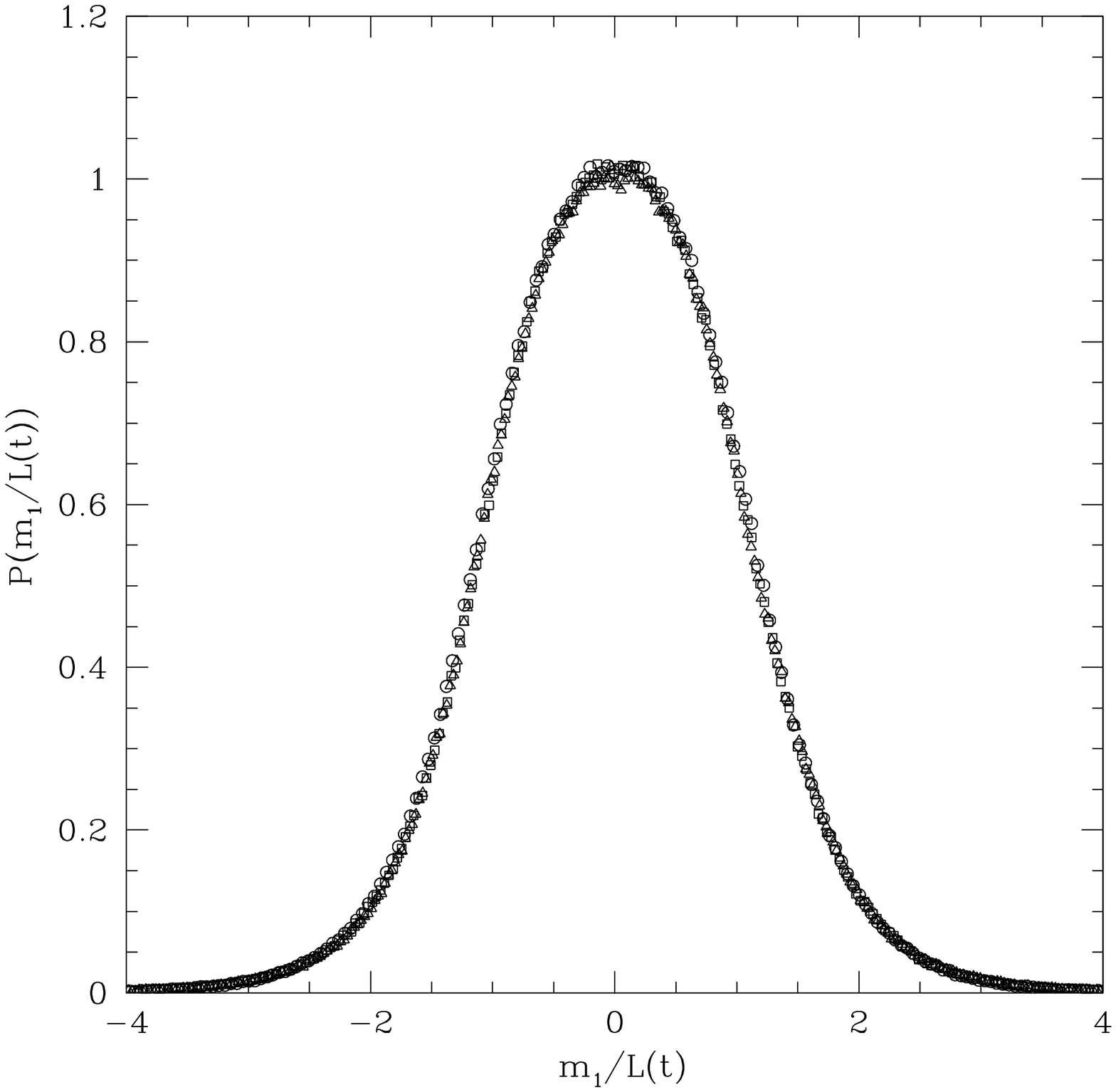,width=8.0cm,height=8.0cm}}
\end{figure}
Fig. 9. Same as Fig.\ 8, but for $g = 1$ for times $\: t=5000\,(\circ),
\:t=10000\,(\Box), \:t=15000\,(\triangle)\:$.

\begin{figure}
\centerline{\psfig{figure=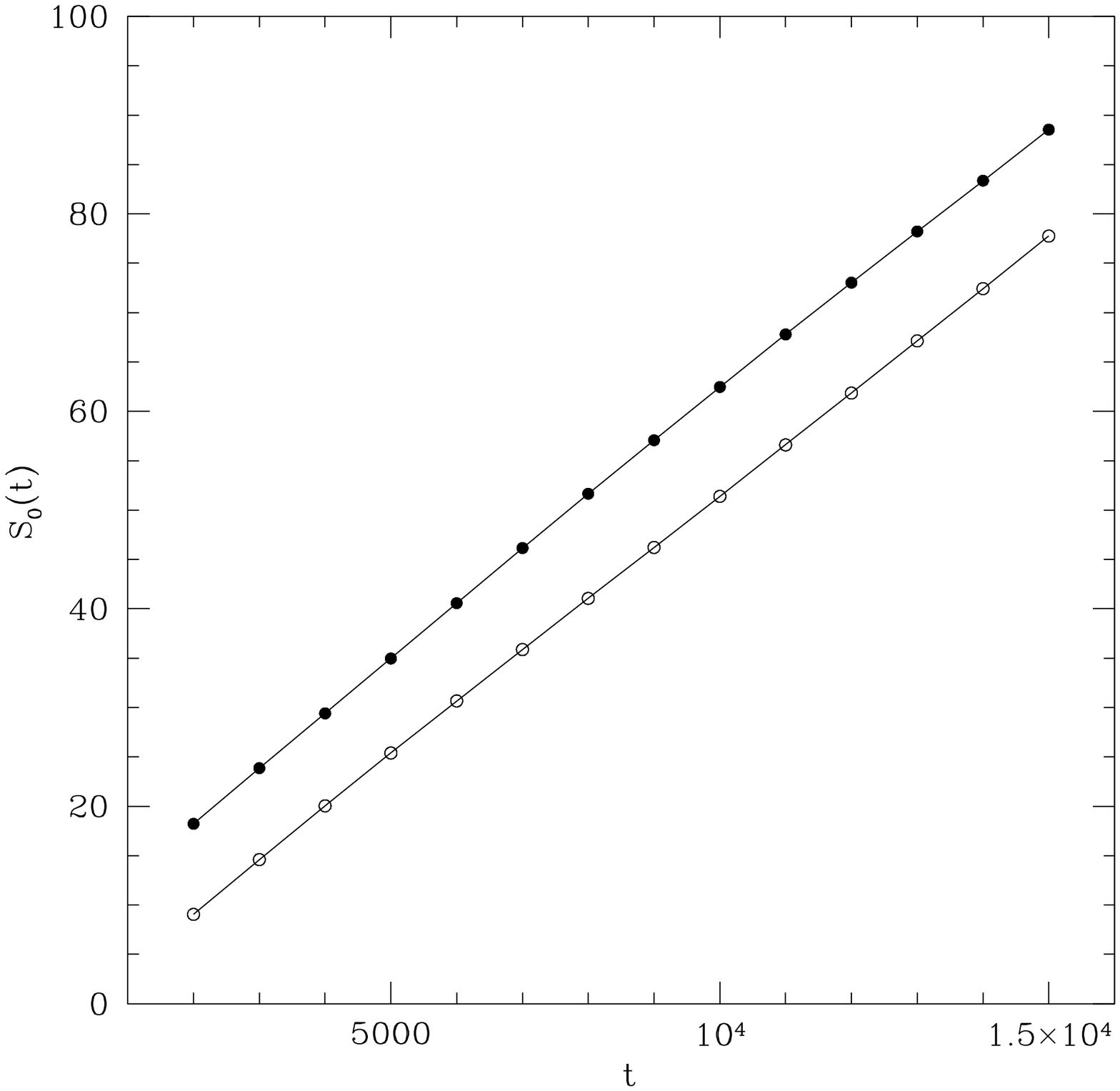,width=8.0cm,height=8.0cm}}
\end{figure}
Fig. 10. Linear growth of $S_0(t)$ with $t$ for $g = 0\,(\circ)$ 
and $g=1\,(\bullet)$.

\begin{figure}
\centerline{\psfig{figure=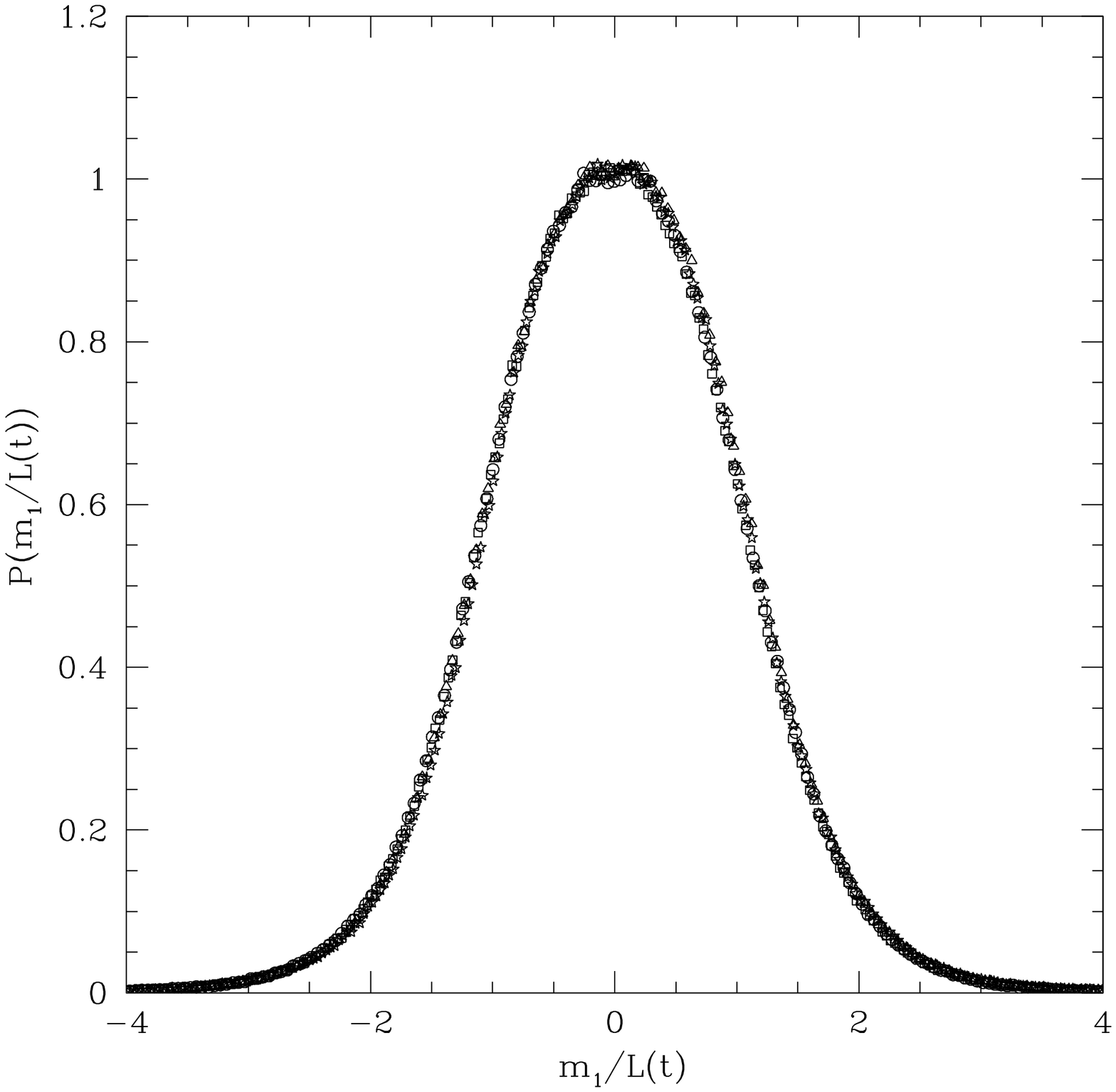,width=8.0cm,height=8.0cm}}
\end{figure} 
Fig. 11. Scaling plot of $P(m_1)$ is independent of $g$.
Data taken for $g = 0 \:( t=5000\,(\circ),\: t=10000\,(\Box) )$ and $g = 1\:(
t=5000\,(\triangle),\: t=10000\,(\star) )$.

\newpage

\begin{figure}
\centerline{\psfig{figure=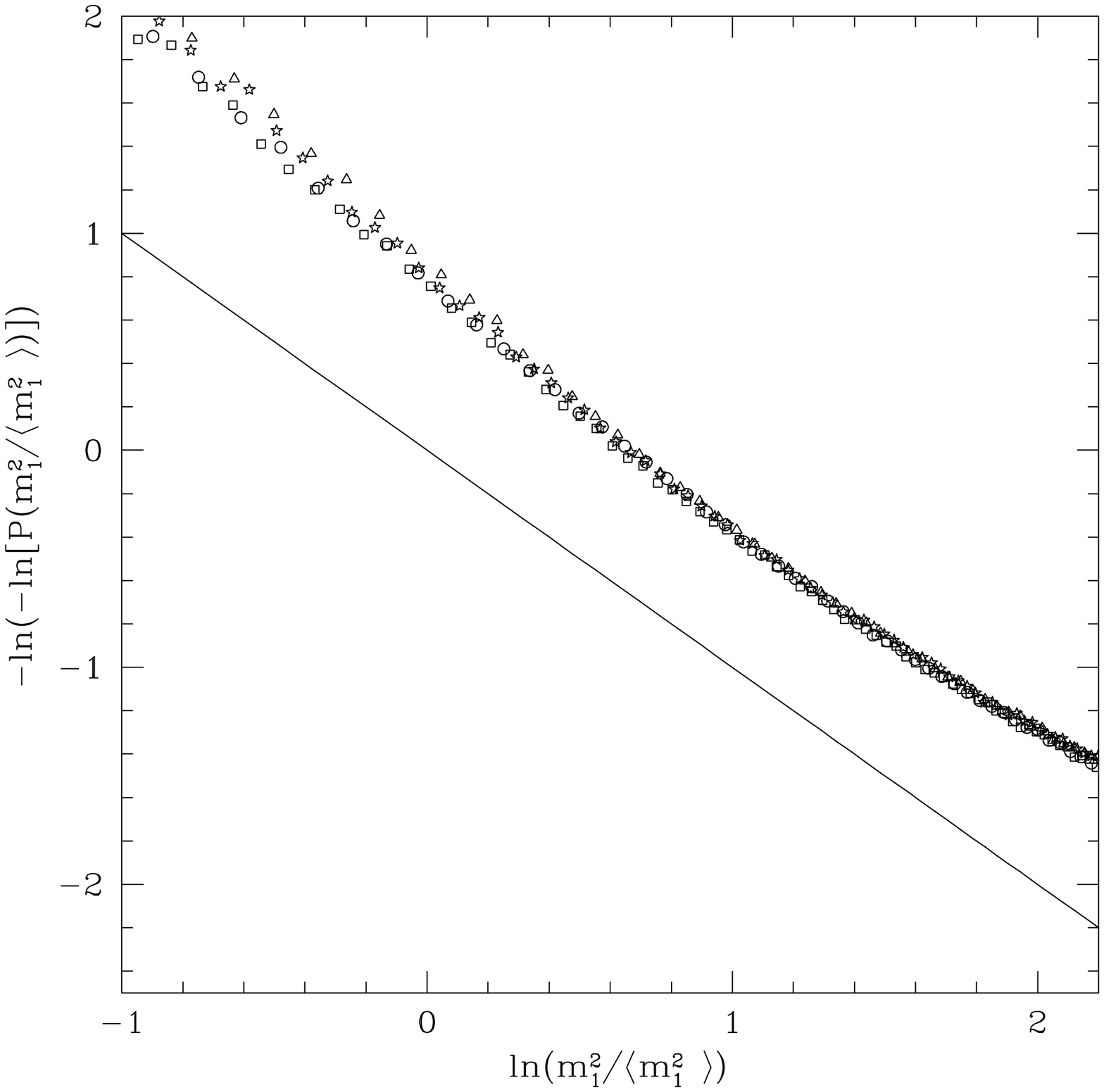,width=8.0cm,height=8.0cm}}
\end{figure}  
Fig. 12. Plot of $ -\ln(\,-\ln\left[P(m^2_1/S_0)
\right]\,)$ versus $\ln(m^2_{1}/S_0)$ for $g = 0\:( t=5000\,(\circ),\:
 t=10000\,(\Box) ) $ and $g = 1 \:(t=5000\,(\triangle),\: t=10000\,(\star) )
 $. The line with slope $-1$ drawn for comparison, highlights the
 deviation of the data from a gaussian at smaller $m_1$. 

\newpage

\begin{figure}
\centerline{\psfig{figure=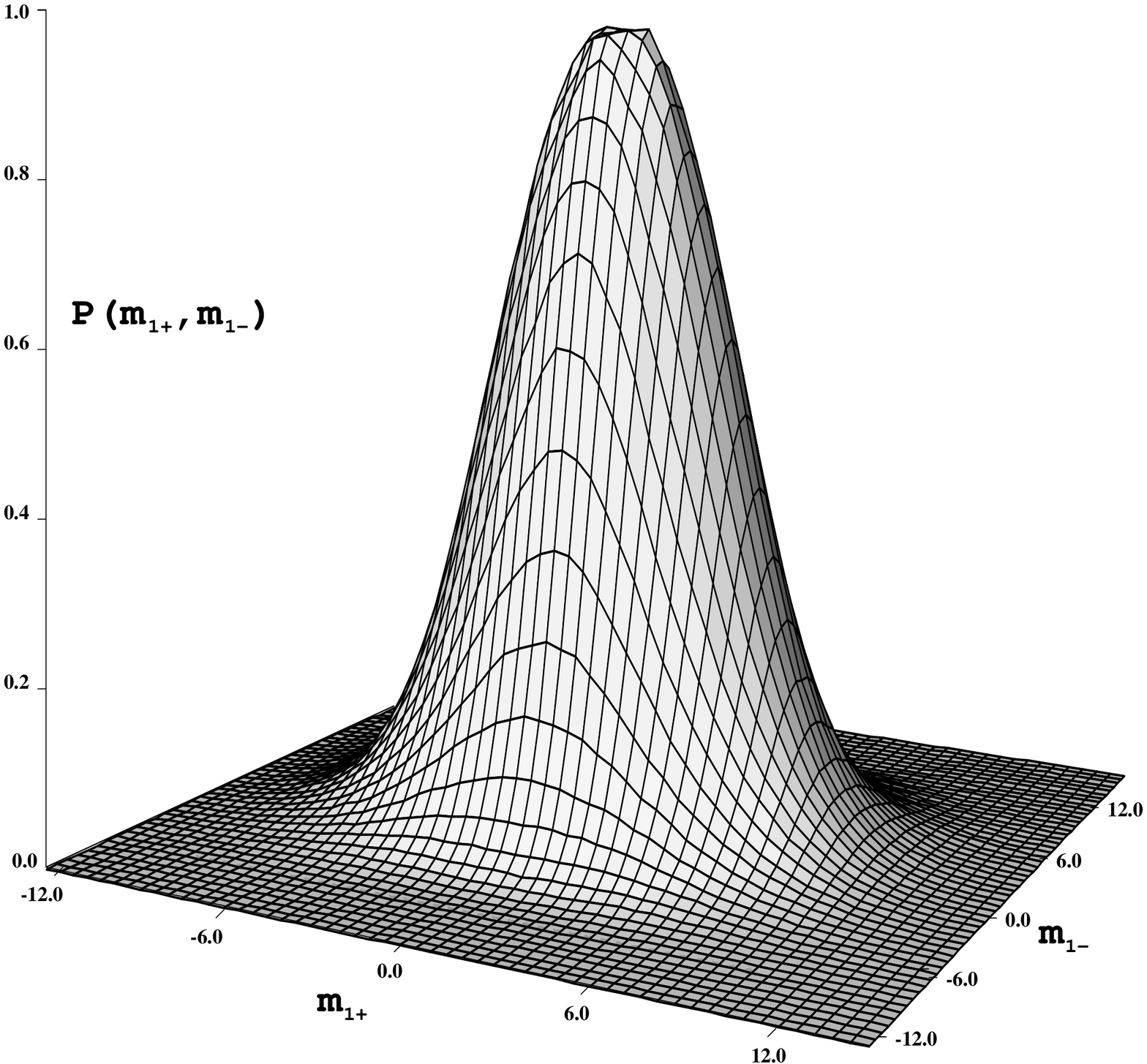,width=8.0cm,height=8.0cm}}
\end{figure} 

Fig. 13. Joint probability distribution 
$P({\vec m}_{+}(12),{\vec m}_{+}(12)) $
for $g = 0$ at $t = 5000$ and $ \vert{\bf r}_1-{\bf r}_2 \vert =
4\sqrt{3} $. 

\newpage

\begin{figure}
\centerline{\psfig{figure=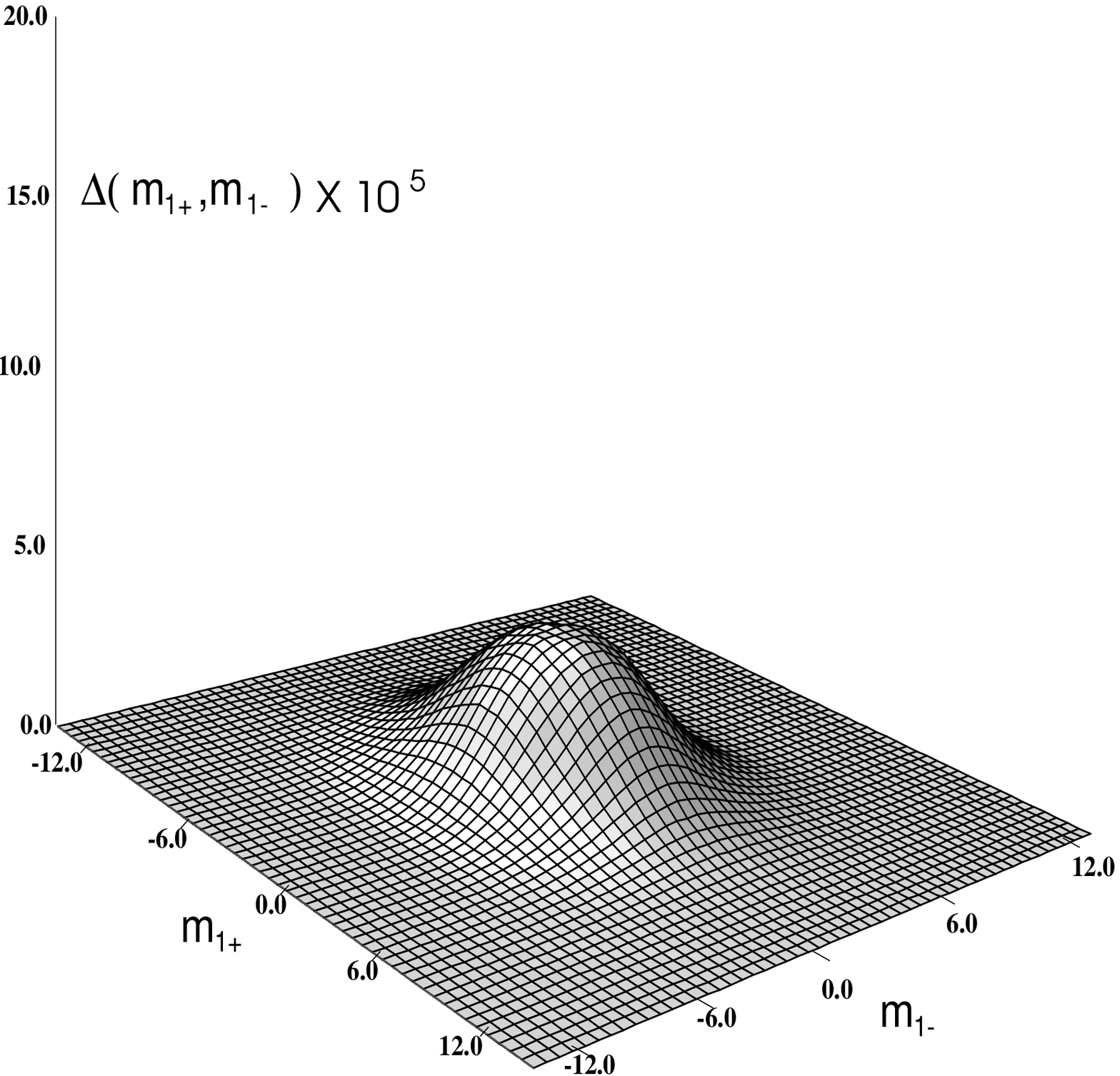,width=8.0cm,height=8.0cm}}
\end{figure} 
Fig. 14. Surface plot of $\Delta(m_{1+}(12), m_{1-}(12))$ (magnified
$10^5$ times) for $ g = 0 $ at $ t = 5000 $ and $\vert{\bf r}_1-{\bf
r}_2 \vert = 4\sqrt{3} $.

\begin{figure}
\centerline{\psfig{figure=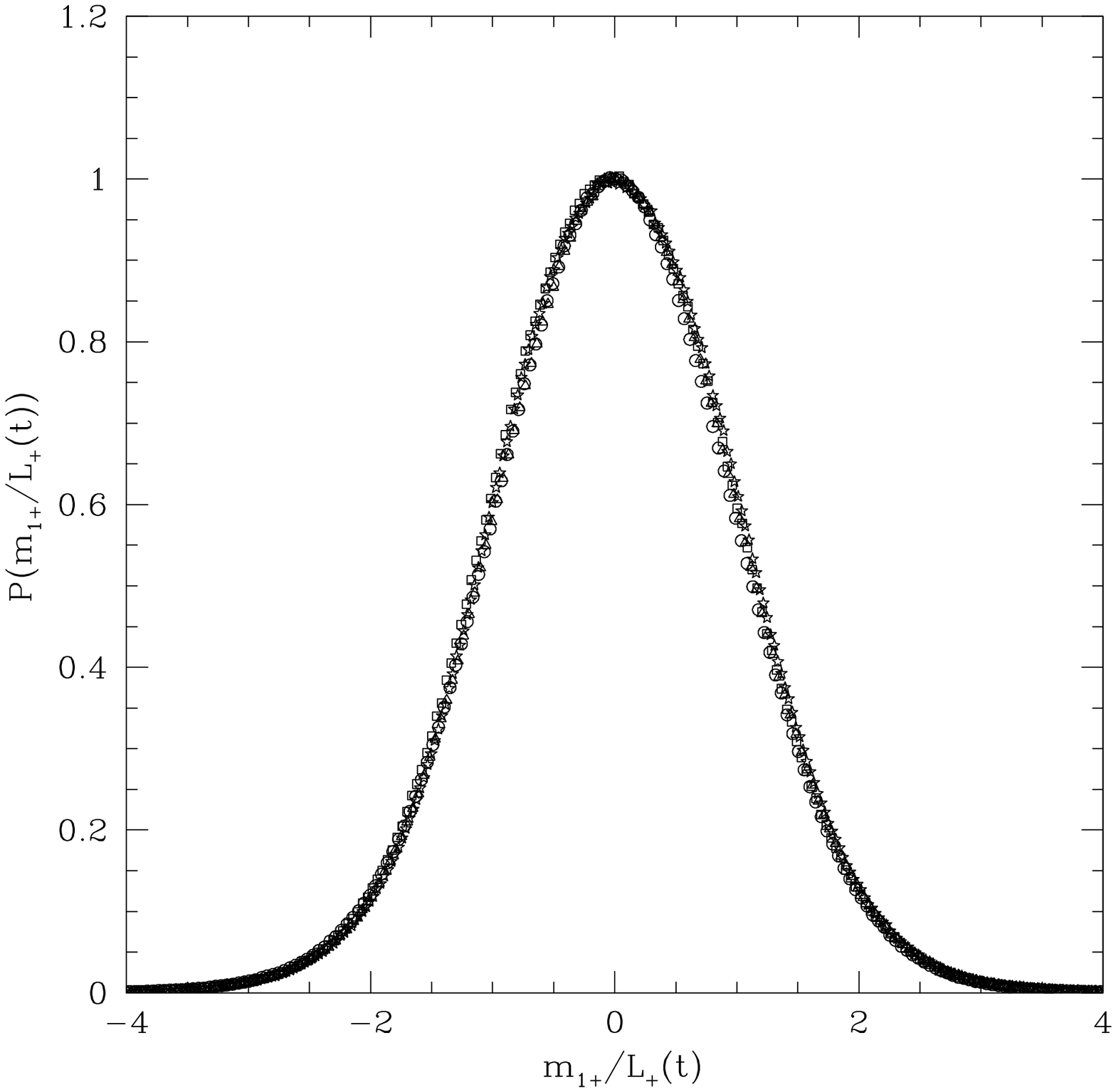,width=8.0cm,height=8.0cm}}
\end{figure}  
Fig. 15. Scaling plot of $P(m_{+1}/L_{+})$, where $L_{+} =
\sqrt{\langle m_{+1}^2({\bf r },t) \rangle}$, for $g = 0 \:(
t=5000\,(\circ),\: t=10000\,(\Box) )$ and $ g=1\:(t=5000\,(\triangle),\:
t=10000\,(\star))$.

\begin{figure}
\centerline{\psfig{figure=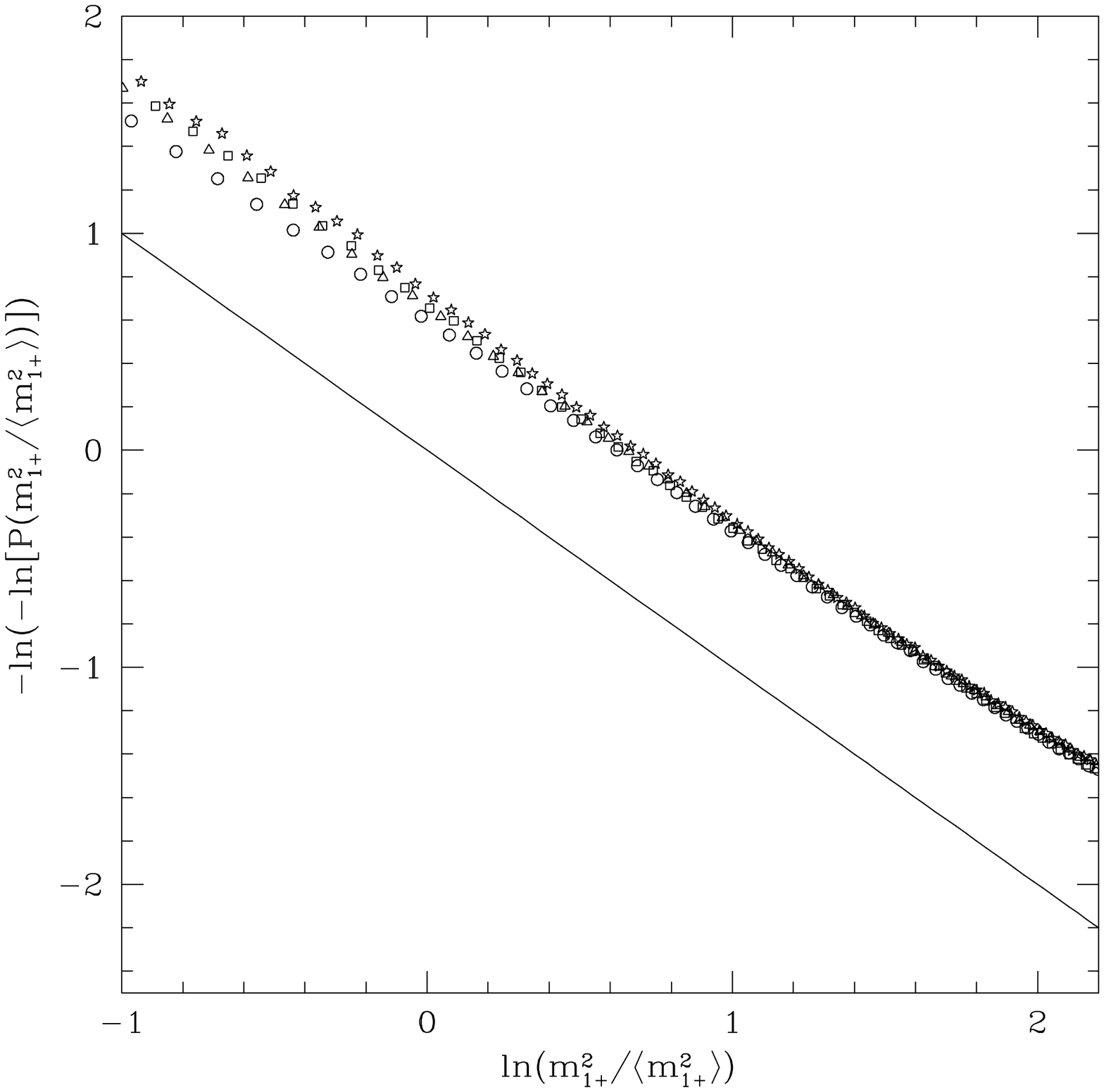,width=8.0cm,height=8.0cm}}
\end{figure} 
Fig. 16. $P(m_{+1}/L_{+})$ shows deviation
from gaussian for small $m_{+}$. Data shown for $g = 0 \:(
t=5000\,(\circ),\: t=10000\,(\Box) )$ and $ g=1\:(t=5000\,(\triangle),\:
t=10000\,(\star))$.

\begin{figure}
\centerline{\psfig{figure=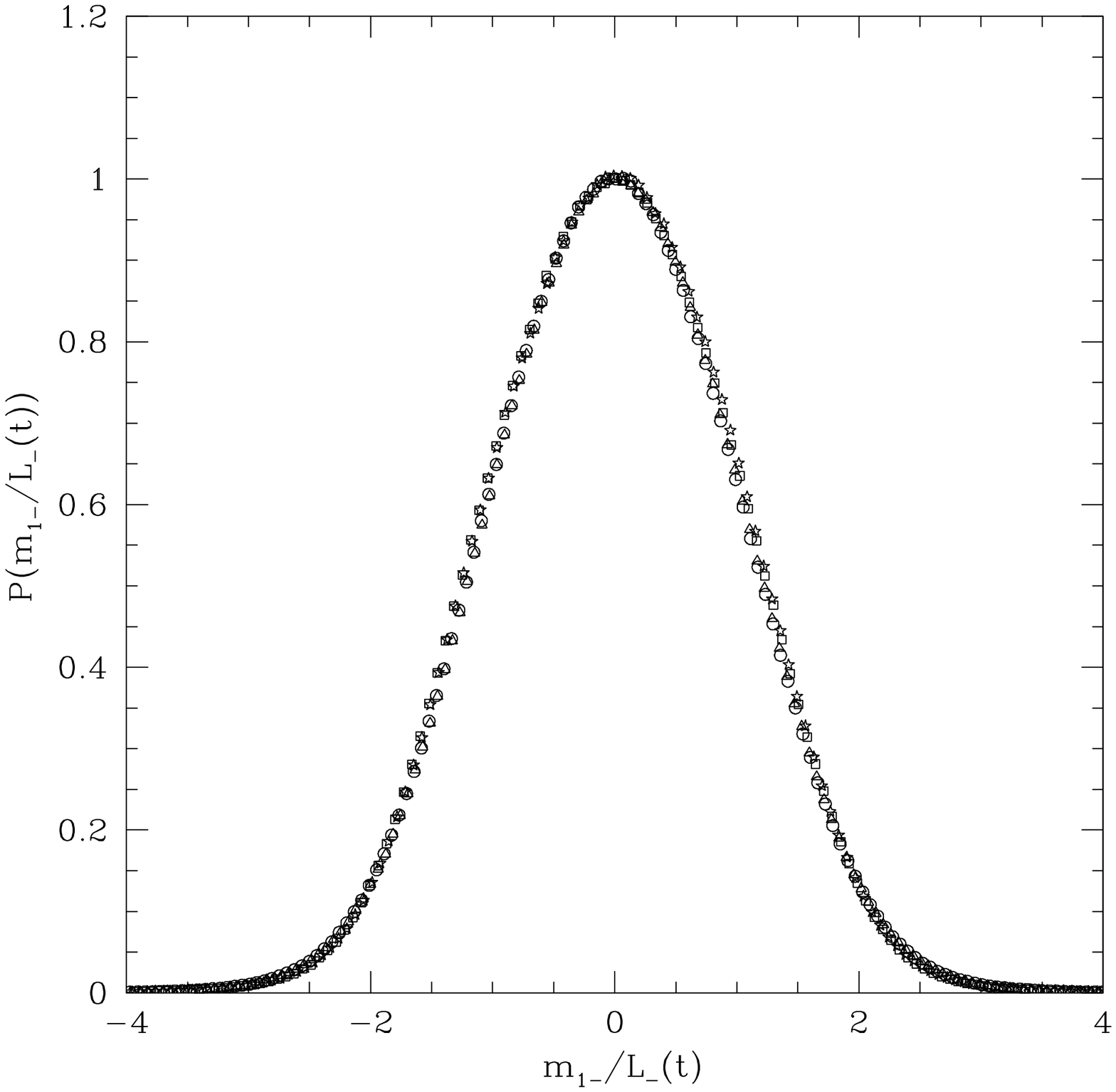,width=8.0cm,height=8.0cm}}
\end{figure}
Fig. 17. Scaling plot of $P(m_{-1}/L_{-})$,where $L_{-} =
\sqrt{\langle m_{-1}^2({\bf r },t) \rangle}$, for $g = 0\:(
t=5000\,(\circ),\: t=10000\,(\Box) ) $ and $ g=1\:(t=5000\,(\triangle),\:
t=10000\,(\star))$.

\begin{figure}
\centerline{\psfig{figure=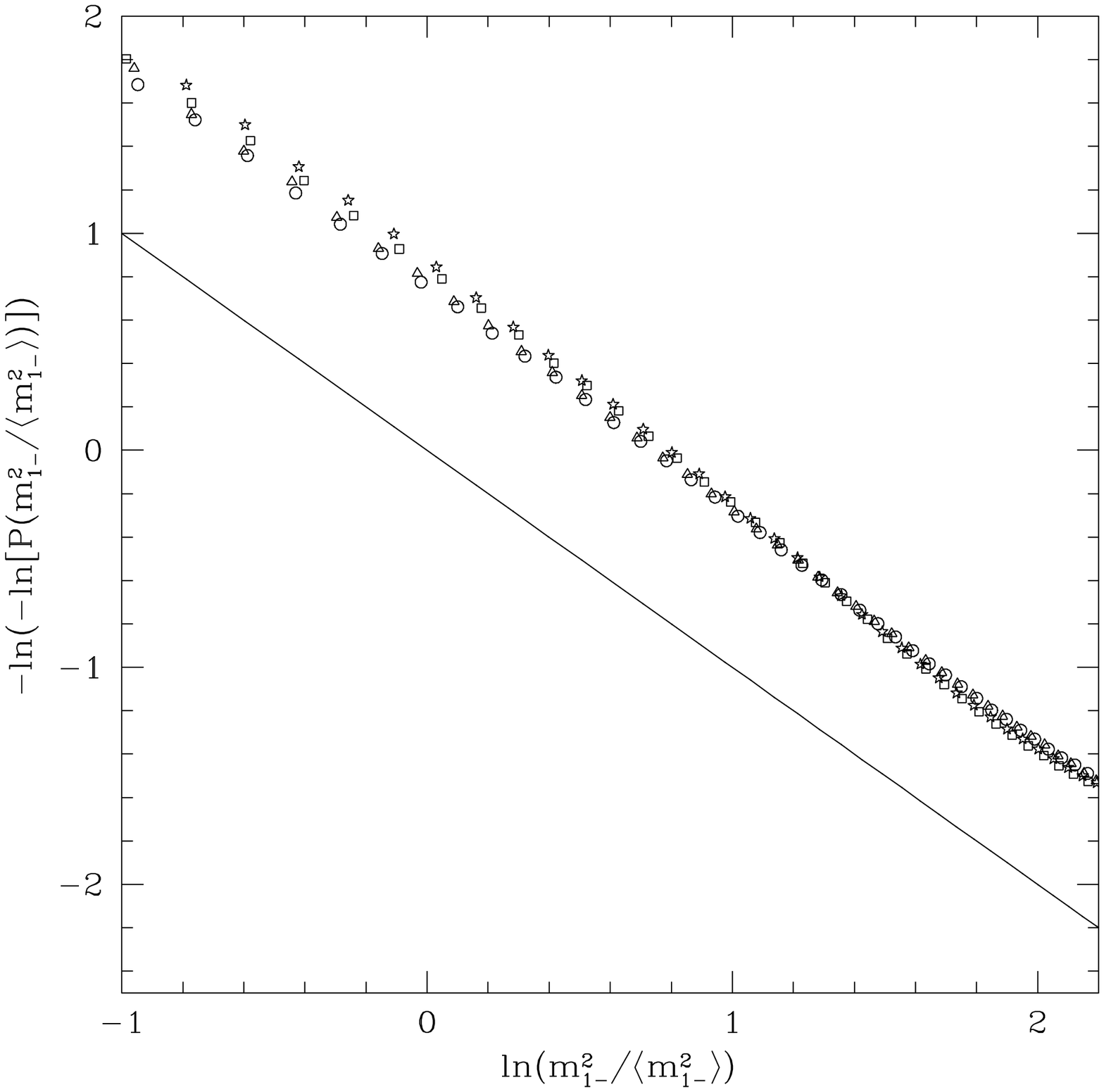,width=8.0cm,height=8.0cm}}
\end{figure}
Fig. 18. $ P(m_{-1}/L_{-})$ shows deviation
from guassian for small $m_{-1}$. Data shown for $g = 0 \:(
t=5000\,(\circ),\: t=10000\,(\Box) )$ and $ g=1\:(t=5000\,(\triangle),\:
t=10000\,(\star))$. 

\begin{references}
\bibitem{NOTE1}
In order to make such a claim, great care should be taken to avoid finite
size effects. The tail of the autocorrelation function gets affected by 
the finite size of the lattice, earlier than the part of $C(r,t)$ used to 
calculate $z$. The decrease in the dip at $g=5$ can only be perceived 
when the system size is large.
\end{references}
\end{document}